 \documentclass[journal,12pt,draftclsnofoot,onecolumn,english]{IEEEtran}
\usepackage{graphicx}
\usepackage{amsmath,amssymb}
\usepackage{cases}
\usepackage{color}
\usepackage{amsfonts}
\usepackage[latin9]{inputenc}
\usepackage{indentfirst}
\usepackage{cite}
\usepackage{caption}
\usepackage{algorithm}
\usepackage{algpseudocode}
\usepackage{booktabs}
\usepackage{blkarray}
\usepackage{subfigure}
\usepackage{multirow}
\usepackage{amsthm}
\usepackage{diagbox}
\makeatletter
\newtheoremstyle{mythm}{3pt}{3pt}{}{16pt}{\bfseries}{:}{.5em}{}
\theoremstyle{mythm}
\newtheorem{theorem}{Theorem}
\newtheorem{example}{Example}
\newtheorem{definition}{Definition}
\newtheorem{remark}{Remark}

\newtheorem{corollary}{Corollary}
\newtheorem{lemma}{Lemma}
\newtheorem{construction}{Construction}

\allowdisplaybreaks[3]

\begin{document}
\title{Hierarchical Cache-Aided Linear Function Retrieval with Security and Privacy Constraints
\author{Yun Kong, Youlong Wu, and Minquan Cheng}
\thanks{Yun Kong and Minquan Cheng are with Guangxi Key Lab of Multi-source Information Mining $\&$ Security, Guangxi Normal University,
Guilin 541004, China.  (e-mail: yunkong2022@outlook.com, chengqinshi@hotmail.com). }
\thanks{Youlong Wu is with the School of Information Science and Technology, ShanghaiTech University,
201210 Shanghai, China. (e-mail: wuyl1@shanghaitech.edu.cn).}
}
\maketitle

\begin{abstract}

The hierarchical caching system where a server connects with multiple mirror sites, each connecting with a distinct set of users, and both the mirror sites and users are equipped with caching memories has been widely studied. However all the existing works focus on single file retrieval, i.e., each user requests one file, and ignore the security and privacy threats in communications. In this paper we investigate  the linear function retrieval problem for hierarchical caching systems with content security and demand privacy, i.e., each user requests a linear combination of files, and meanwhile the files in the library are protected against wiretappers and users' demands are kept unknown to other users and unconnected mirror sites. First we propose a new combination structure named hierarchical placement delivery array (HPDA), which characterizes the data placement and delivery strategy of a coded caching scheme. Then we construct two classes of HPDAs. Consequently two classes of schemes with or without security and privacy are obtained respectively where the first dedicates to minimizing the transmission load for the first hop and can achieve the optimal transmission load for the first hop if ignoring  the security and privacy constraints; the second has more flexible parameters on the memory sizes and  a lower subpacketization compared with the first one, and   achieves a tradeoff between subpacketization and transmission loads.
\end{abstract}

\section{Introduction}
Caching system is an efficient way to reduce the transmission during the peak traffic hours by shifting traffic from peak to off peak hours.
Maddah-Ali and Niesen in \cite{MN} showed that the transmission load during the peak traffic hours can be further reduced by utilizing the cached contents to generate additional multicast opportunities among the users. In the original $(K,M,N)$ caching system, a single server having a library with $N$ files of equal size is connected to $K$ cache-aided users who can cache contents of size at most $M$ files through an error-free shared-link. An $F$-division $(K,M,N)$ coded caching scheme contains two phases, i.e., placement phase during the off peak hours and delivery phase during the peak hours. In placement phase, the server divides each file into $F$ packets with equal size and then places at most $MF$ packets into each user's cache without any information about users' demands. $F$ is referred as the subpacketization. Since the packets are directly cached by the users, this placement strategy is called uncoded placement. In delivery phase, each user requests a file from the server randomly and the server broadcasts some coded messages with size of at most $R$ files to users such that all the users can decode their requested files with the help of their cached packets. The value of $R$ is referred as transmission load.

Maddah-Ali and Niesen proposed the first centralized coded caching scheme \cite{MN} (MN scheme) and the first decentralized coded caching scheme \cite{MND} (MN decentralized scheme) respectively. It is worth noting that when $N\geq K$, MN scheme has the minimum transmission load under uncoded placement in \cite{WTDPP}.  However, the subpacketization $F$ of the MN scheme increases exponentially with the growing on user number $K$. In order to design a scheme with low subpacketization, the authors in \cite{YCTC} proposed a combination structure named placement delivery array (PDA) to simultaneously characterize the placement and delivery phase of a coded caching scheme. The MN scheme could also be represented by a special PDA which is referred to as MN PDA. It is worth noting that the MN scheme has the minimum subpacketization for the fixed minimum transmission load among all the schemes which can be realized by PDAs \cite{CJTY}. Besides PDA, there also exists some other combination structures to characterize a coded caching scheme, such as \cite{SCZYG,YTCC,CLTW}, etc.

\begin{figure}[http!]
\centering
\includegraphics[scale=0.35]{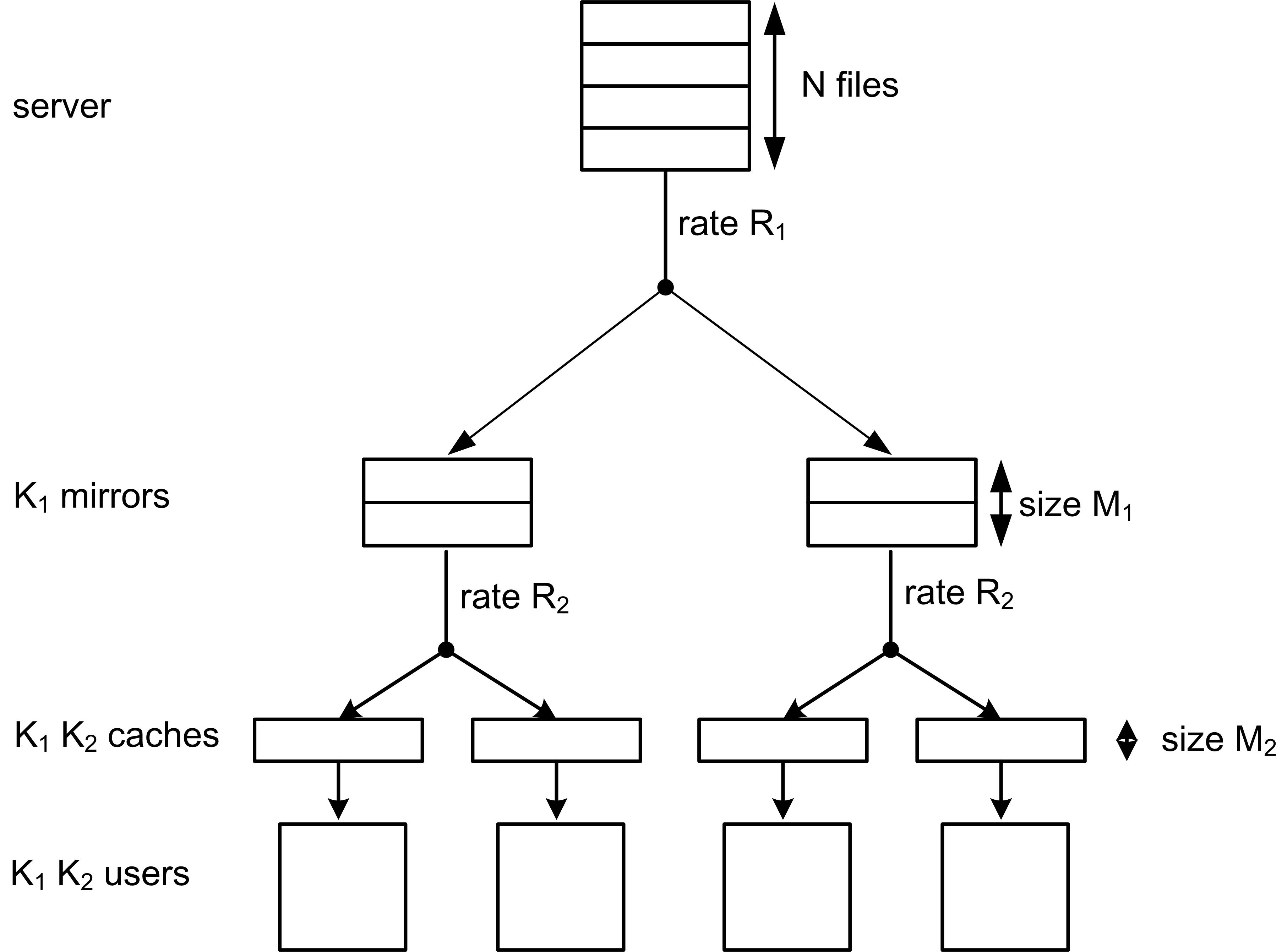}
\caption{The $(K_1,K_2;M_1,M_2;N)$ hierarchical  caching system with $N=4$, $K_1=K_2=2$, $M_1=2$ and $M_2=1$.}
\label{fig-model}
\end{figure}
In reality, the caching systems consist of multiple layers in most cases, i.e., between the central server and the end users, there are some in-between devices, and all these devices are arranged in a tree-like hierarchy with the central server at the root node while the end users act as the leaf nodes. Between each layer, the parent node communicates with its children nodes. As illustrated in Fig. \ref{fig-model}, a $(K_1$, $K_2$; $M_1$, $M_2$; $N)$ hierarchical caching system was first studied in \cite{KNMD}. That is, a two-layer hierarchical network consists of a single origin server and $K_1$ cache-aided mirror sites and $K_1K_2$ cache-aided users where the server hosts a collection of $N\geq K_1K_2$ files with equal size, each mirror site and each user have memories of size $M_1$ files and $M_2$ files respectively where $M_1$, $M_2\leq N$. The server is connected through an error-free shared link to each mirror site. Each mirror site is connected through an error-free broadcast link to $K_2$ users and each user is connected to only one mirror site. The objective is to design a scheme with the transmission loads $R_1$ and $R_2$ for the first layer (from sever to mirror sites) and the second layer (from each mirror site to its connected users) and subpacketization as small as possible. Under the uncoded data placement, define the optimal transmission loads for the first layer and second layer as $R^*_1$, $R^*_2$ respectively.

There are some studies on designing coded caching schemes for the hierarchical network \cite{KNMD,ZWXWL,WWCY}, and all the existing works are performed under a two-subsystem model which is derived by dividing the $(K_1,K_2;M_1,M_2;N)$ hierarchical caching system into two subsystems. The first one includes the sever containing $N$ files each of which has a $\alpha$ fraction of each original file, $K_1$ mirror sites with their whole cache memory size $M_1$, and $K_1K_2$ users with their $\beta$ fraction of memory size $M_2$; the second includes a server containing $N$ files each of which has the left $1-\alpha$ fraction of each file and $K_1K_2$ users with their $1-\beta$ fraction of memory size $M_2$. Clearly, the second subsystem ignores mirror sites' caching abilities, which can be regarded as a $(K_1K_2,\frac{1-\beta}{1-\alpha} M_2,(1-\alpha))N)$ original caching system, thus all the existing works mainly focus on designing schemes for the first subsystem. In \cite{KNMD}, the authors directly used the MN decentralized scheme in each layer. Then the requested files are built in each layer, so it results in a high transmission load of $R_1$. This scheme is referred as KNMD scheme. By extending KNMD scheme to the case with centralized data placement \cite{MN},  the transmission loads of the first and second layer, denoted by $R^\text{KNMD}_1$ and $R^\text{KNMD}_2$ are given by
\begin{eqnarray}
\label{DHCC-B}
\begin{split}
R^\text{KNMD}_1(\alpha,\beta)&\triangleq   \alpha\cdot K_2\cdot r_c\left(\frac{M_1}{\alpha N},K_1\right)
+(1-\alpha)\cdot r_c\left(\frac{(1-\beta)M_2}{(1-\alpha)N},K_1K_2 \right),&\\
R^\text{KNMD}_2(\alpha,\beta)&\triangleq  \alpha\cdot r_c\left(\frac{\beta M_2}{\alpha N},K_2\right)
+(1-\alpha)\cdot r_c\left(\frac{(1-\beta)M_2}{(1-\alpha)N},K_2 \right),
\end{split}
\end{eqnarray} for some $\alpha$ and $\beta$, where $r_c\left(\frac{M}{N},K\right)  \triangleq \frac{K (1-M/N)}{1+KM/N}$ is the transmission load of the $(K,M,N)$ MN scheme for any memory ration $\frac{M}{N}\in\{0,\frac{1}{K},\frac{2}{K},\ldots,1\}$. Without building the whole file in each layer, the authors in \cite{WWCY} proposed an improved scheme (the WWCY scheme) by utilizing two MN schemes. The WWCY scheme achieves the minimum load of $R_2$ under uncoded placement, and the load of the first layer $R_1$ is smaller than the KNMD scheme, where the transmission loads of the first and second layer, denoted by $R^\text{WWCY}_1$ and $R^\text{WWCY}_2$,  are 
\begin{eqnarray}
\label{eq-load-W}
\begin{split}
R^\text{WWCY}_1(\alpha,\beta)&\triangleq   \alpha \cdot r_c\left(\frac{M_1}{\alpha N},K_1\right)r_c\left(\frac{\beta M_2}{\alpha N},K_2\right)
+(1-\alpha)r_c\left(\frac{(1-\beta)M_2}{(1-\alpha)N},K_1K_2\right),\\
 R^\text{WWCY}_2(\alpha,\beta)&\triangleq  \alpha\cdot r_c\left(\frac{\beta M_2}{\alpha N},K_2\right)
+(1-\alpha)\cdot r_c\left(\frac{(1-\beta)M_2}{(1-\alpha)N},K_2 \right).
\end{split}
\end{eqnarray}

Note that under uncoded placement, schemes in \cite{KNMD} and \cite[Section VI]{WWCY} both achieve the optimal transmission load of the second layer  when $\alpha=\beta$, i.e.,
$R^*_2=r_c\left(\frac{  M_2}{  N},K_2\right).$
However, the question what is the optimal transmission load of the first layer is still unknown, and the schemes in \cite{KNMD,ZWXWL,WWCY} utilize the MN scheme or MN decentralized scheme, whose subpacketization increases exponentially with the growing on user number.

In practical scenarios, the content security against the wiretapping and users' demand privacy are also paramount aspects which have been studied in many caching systems such as the original caching system \cite{SATRCT,YDT,WGC,GCJSBKN}, combination networks \cite{ZAYA}, D2D network \cite{WSJTC,AZSA}, multiaccess network \cite{WCLGC}. The secure delivery considers that the files in the library are kept secret from any wiretapper which can overhear from the shared-link. The proposed scheme in \cite{SATRCT} uses security keys known by both server and users to encrypt all the transmitted messages, and   the cost is negligible when the number of users and files are large. The demand privacy considers that each users' demand is kept unknown to other users. The proposed schemes in \cite{WGC,GCJSBKN} ensure the demand privacy by adding $N(K-1)$ virtual users and hide the real $K$ users' demands among the total $NK$ users. The authors in \cite{YDT} used the i.i.d. privacy vectors allocated by the server to guarantee the demand privacy against colluding users, and the demand privacy against colluding users was first proposed in \cite{WSJTC} for a D2D network, where the colluding means that some user share their cached contents and demands. Recall that   \cite{YDT} also considered the private and secure delivery  by the superposition of security keys and privacy vectors based on PDA. However, for security in hierarchical network, the only work \cite{ZBVBKT} proposed a secure coded caching scheme by combining the security key scheme in \cite{SATRCT} with KNMD scheme, and for demand privacy in hierarchical network there is no existing works.

In addition, all the above mentioned  coded caching schemes are performed under the case that each user only requests one file (i.e., single file retrieval). The author in \cite{WSJTDC} proposed a general requesting formulation called linear function retrieval, i.e., each user requests a linear combination of files, which has been studied in \cite{YDT,MYTD,YTD}. Obviously the single file retrieval is a special case of linear function retrieval.

\subsection{Contribution and Paper Organization}
\label{subsec:con-org}
In this paper we focus on designing coded caching schemes with or without security and privacy for linear function retrieval problem in the hierarchical caching system where $N>K_1K_2$, and consider reducing the transmission loads $R_1$, $R_2$ or subpacketization. The main result are as follows.
\begin{itemize}
\item Inspired by the concept of PDA, we propose a new combination structure referred to as hierarchical placement delivery array (HPDA), which could be used to characterize the placement and delivery phases of a hierarchical coded caching scheme. So designing a hierarchical coded caching scheme is transformed to constructing an appropriate HPDA, and the schemes in \cite{ZWXWL,WWCY} also can be realized from the point of view of HPDA.
\item Based on the    HPDA, we present how to design hierarchical coded caching schemes for linear function retrieval with content security and demand  privacy, i.e., each user requests a linear combination of files, and meanwhile the files in the library are protected against wiretappers and users' demands are kept unknown to other users and unconnected mirror sites. The proposed scheme  achieves the same transmission loads   and subpacketization as the basic scheme (where each user requests a single file without security and privacy), with a negligible cost of cache memories especially for the large file number. This indicates that    the privacy and security can be achieved  without increasing communication loads for hierarchical  linear function retrieval problem.
\item For a given PDA, by a grouping method we can obtain a class of HPDA which leads to a coded caching scheme dedicating to minimizing the first layer transmission load $R_1$. When the given PDA is MN PDA, the corresponding scheme without security and privacy achieves the optimal transmission load $R_1$ under uncoded data placement. However, the grouping method only works for some certain circumstances, and for the general case we have the following hybrid construction, which has a flexible choice of mirror sites' number, user number and memory ratios.
\item We provide a hybrid construction of HPDA based on any two PDAs, leading to a new    scheme with flexible subpacketiztion when we choose the basic PDAs with various subpacketiziations. This  new scheme achieves a flexible tradeoff between the subpacketization and transmission loads. In addition if the basic PDAs are MN PDAs and each user requests a single file without security and privacy, then the scheme realized by our HPDA is exactly the WWCY scheme  in \cite{WWCY}.
\end{itemize}

The rest of this paper is organized as follows. The hierarchical caching model is introduced in Section \ref{sec:System}. We review PDA and introduce the structure of HPDA in Section \ref{sec:HPDA}. The main results and some intuitions about constructions of HPDA and performance analysis are listed in Section \ref{sec:main-result}. The coded caching scheme with or without security and privacy is proposed in Sections \ref{sec:proof-th-1}. The proofs of our main results are given in Sections \ref{sec:proof-th-2}, \ref{sec:proof-th-3} and Appendix. Finally we conclude this paper in Section \ref{sec:conclusion}.
\subsection{Notations}
 The following notations are used in this paper.

\begin{itemize}

  \item For any positive integers $a$ and $b$ with $a<b$, let $[a:b]\triangleq\{a,a+1,\ldots,b\}$, $[a:b)\triangleq\{a,a+1,\ldots,b-1\}$ and $[a]\triangleq\{1,2,\ldots,a\}$. Let $\binom{[a]}{t}\triangleq \{\mathcal{V}\ |\ \mathcal{V}\subseteq [a],|\mathcal{V}|=t\}$ for any positive integers $t\leq a$.
  \item Given an array $\mathbf{P}=(p_{j,k})_{j\in[F], k\in[K]}$ with alphabet $[S]\cup \{*\}$, we define $\mathbf{P}+a=(p_{j,k}+a)_{j\in[F], k\in[K]}$ and $\mathbf{P}\times a=(p_{j,k}\times a)_{j\in[F], k\in[K]}$  for any integer $a$, where $a+*=*, a\times *=*$.
  \item For a positive integer $n$, $\mathbb{F}^{n}_q$ is the n dimensional vector space over the field $\mathbb{F}_q$.
  \item For any two integer set $\mathcal{T}_1$, $\mathcal{T}_2$, define $\mathcal{T}_1\times\mathcal{T}_2=\{(t_1,t_2)\ |\ t_1\in\mathcal{T}_1,\ t_2\in\mathcal{T}_2\}$.
\end{itemize}

\section{Problem Definitions}
\label{sec:System}
In this section, we   describe the hierarchical coded caching problem with linear functions retrieval under security and privacy constraints.

The network model is shown in    Fig. \ref{fig-model} which consists of a single server, $K_1$ mirror sites and $K_1K_2$ users. The server connects to  $K_1$ mirror sites via an error-free shared link,  and each mirror site connects with $K_2$ users via another error-free shared link. Denote the $k_1$-th mirror site by ${\text M}_{k_1}$ and the $k_2$-th user attached to mirror site ${\text M}_{k_1}$ as $\text{U}_{k_1,k_2}$, for  $k_1\in[K_1]$, $k_2\in [K_2]$, and the set of users attached to the $k_1$-th mirror site as $\mathcal{U}_{k_1}$. For any $\mathcal{T}_1\subseteq[K_1]$ and $\mathcal{T}_2\subseteq[K_2]$, define $\mathcal{M}_{\mathcal{T}_1}\triangleq\{\text{M}_{k_1}| \ k_1\in\mathcal{T}_1\}$ and $\mathcal{U}_{\mathcal{T}_1,\mathcal{T}_2}\triangleq\{\text{U}_{k_1,k_2}| \ k_1\in\mathcal{T}_1, k_2\in\mathcal{T}_2\}$.


We assume that the server contains a collection of $N$ files, denoted by $\mathcal{W} = \{W_1, W_2,\ldots,W_{N}\}$, each of which is uniformly distributed over $\mathbb{F}^B_{q}$, where $q$ is the prime power and $B$ is the length of the file; each mirror site and user has memory size of $M_1$ and $M_2$ files, respectively, for some $M_1$, $M_2\geq 0$; the server privately generates a random variable $P$ from some probability space $\Omega$ to protect the security and privacy of the caching system. A $(K_1,K_2;M_1,M_2;N)$ caching system with privacy and security contains two phases.

\subsubsection{Placement phase} 
     Each mirror site ${\text M}_{k_1}$, for $k_1\in [K_1]$, caches some parts of the file by a cache function  $\varphi_{k_1}:\Omega\times \mathbb{F}^{NB}_{q}\mapsto\mathbb{F}^{M_1B}_{q}$. Then the cache contents of the mirror site ${\text M}_{k_1}$ can be written as
     \begin{eqnarray*}
      \mathcal{Z}_{k_1}=\varphi_{k_1}(P,\mathcal{W}),\ \  \forall k_1 \in [K_1].
    \end{eqnarray*}
     For any $\mathcal{T}_1\subseteq[K_1]$, define $\mathcal{Z}_{\mathcal{T}_1}\triangleq\{\mathcal{Z}_{k_1}| \ k_1\in\mathcal{T}_1\}$.

     Each user $\text{U}_{k_1,k_2}$,  for  $k_1\in[K_1]$ and $k_2\in [K_2]$,  caches some parts of the files by a cache function $\phi_{k_1,k_2}:\Omega\times \mathbb{F}^{NB}_{q}\mapsto\mathbb{F}^{M_2B}_{q}$. Then the cache contents of user $\text{U}_{k_1,k_2}$ can be written as
    \begin{eqnarray*}
      \widetilde{\mathcal{Z}}_{k_1,k_2}=\phi_{k_1,k_2}(P,\mathcal{W}),\ \  \forall k_1 \in [K_1],\ \forall k_2 \in [K_2].
    \end{eqnarray*}
    For any $\mathcal{T}_1\subseteq[K_1]$ and  $\mathcal{T}_2\subseteq[K_2]$, define $\widetilde{\mathcal{Z}}_{k_1,\mathcal{T}_2}\triangleq\{\widetilde{\mathcal{Z}}_{k_1,k_2}| \ k_2\in\mathcal{T}_2\}$, $\widetilde{\mathcal{Z}}_{\mathcal{T}_1,\mathcal{T}_2}\triangleq\{\widetilde{\mathcal{Z}}_{k_1,k_2}| \ k_1\in\mathcal{T}_1, k_2\in\mathcal{T}_2\}$.

    The $M_1$ and $M_2$ are the storage capacities at each mirror site and user, respectively, for $M_1,M_2\in[0:N]$.
\subsubsection{Delivery phase} Assume that each user $\text{U}_{k_1,k_2}$ wants to retrieve the element-wise linear combination of $N$ files, defined  as  $$L_{\mathbf{d}_{k_1,k_2}}\triangleq d^{(1)}_{k_1,k_2}W_1+\ldots+d^{(N)}_{k_1,k_2}W_N.$$ Then the request vector from user can be represented as $\mathbf{d}_{k_1,k_2}=(d^{(1)}_{k_1,k_2},d^{(2)}_{k_1,k_2},$ $\ldots,d^{(N)}_{k_1,k_2})\in\mathbb{F}^{N}_q$.
We denote the request vectors from the users served by mirror site $\text{M}_{k_1}$, i.e., the users in $\mathcal{U}_{k_1}$, by
\begin{eqnarray*}
\mathbf{D}_{k_1}=\left(\small{
     \begin{array}{c}
       \mathbf{d}_{k_1,1} \\
       \vdots  \\
       \mathbf{d}_{k_1,K_2}  \\
     \end{array}}
   \right),
\end{eqnarray*}
and all the request vectors can be represented by a demand matrix
\begin{eqnarray*}
\mathbf{D}=\left(\small{
     \begin{array}{c}
       \mathbf{D}_{1} \\
       \vdots  \\
       \mathbf{D}_{K_1}  \\
     \end{array}}
   \right).
   \end{eqnarray*}
For any $\mathcal{T}_1\subseteq[K_1]$, $\mathcal{T}_2\subseteq[K_2]$, define $\mathcal{D}_{\mathcal{T}_1,\mathcal{T}_2}\triangleq\{\mathbf{d}_{k_1,k_2}| \ k_1\in\mathcal{T}_1, k_2\in\mathcal{T}_2\}$. Let    $\mathcal{D}_{\mathcal{T}_1,[K_2]}$ and $\mathcal{D}_{[K_1]}$ be shortened by $\mathcal{D}_{\mathcal{T}_1}$ and  $\mathcal{D}$, respectively. From our assumption, the demand matrix, the files and the random variables $P$ are independent, then we have
     \begin{eqnarray*}
       H(\mathcal{D},\ P,\ \mathcal{W}) &=& \sum\limits_{k_1=1}\limits^{K_1}\sum\limits_{k_2=1}\limits^{K_2}H(\mathbf{d}_{k_1,k_2})+H(P)+\sum\limits_{n=1}\limits^{N}H(W_n).
     \end{eqnarray*}

There are two types of message transmitting over the network.
    \begin{itemize}
    \item \textbf{The messages sent by the server over the first layer:} Based on the contents cached by mirror sites and users and the demand matrix, the server generates a coded signal $X^{\text{server}}$ by an encoding function $\chi:\Omega\times\mathbb{F}^{K_1K_2N}_q\times\mathbb{F}^{NB}_q\mapsto\mathbb{F}^{R_1B}_q$, i.e., $X^{\text{server}} = \chi(P,\mathcal{D},\mathcal{W})$. Here $R_1$ is the transmission load of the first layer from server to each mirror site. For the security of the first layer, the server does not want any wiretapper to obtain any information about the files in the library and users' requests from $X^{\text{server}}$. Then the  security constraint for the first layer is defined as follows:
        \begin{eqnarray}
            \label{Security I}
            &\text{[Security I]}& \ \ \ I(\mathcal{D}, \mathcal{W};\ X^{\text{server}})=0.
        \end{eqnarray}
     In addition, the following colluding attack from the mirror site is considered. For any subset $\mathcal{T}_1\subseteq [K_1]$, each mirror site in $\mathcal{M}_{\mathcal{T}_1}$ can not obtain any information about the request vectors of the other users in $\mathcal{U}_{[K_1]\setminus \mathcal{T}_1,[K_2]}$, i.e., 
        \begin{eqnarray}
            \label{Privacy I}
            &\text{[Privacy I]}& \ \ \ I(\mathcal{D}_{[K_1]\backslash \mathcal{T}_1};\ X^{\text{server}},\ \mathcal{D}_{\mathcal{T}_1},\ \mathcal{Z}_{\mathcal{T}_1},\ \mathcal{W})=0, \ \ \ \forall \mathcal{T}_1\subseteq [K_1].
        \end{eqnarray}

    \item \textbf{The messages sent by mirror site over the second layer:} Based on the messages sent by the server, the contents cached by mirror sites and users and the demand matrix, each mirror site $\text{M}_{k_1}$ generates a coded signal $X^{\text{mirror}}_{k_1}$ by an encoding function $\kappa:\Omega\times\mathbb{F}^{K_2N}_q\times\mathbb{F}^{M_2B}_q\times X^{\text{server}}\mapsto\mathbb{F}^{R_2B}_q$, i.e., $X^{\text{mirror}}_{k_1} = \kappa(P,\mathbf{D}_{k_1},\mathcal{Z}_{k_1},X^{\text{server}})$.
        Here $R_2$ is the transmission load of the second layer from a mirror site to its attached users.

        For the security of the second layer, the server does not want any wiretapper to obtain any information about the files in the library and users' requests from the coded signal $X^{\text{mirror}}_{k_1}$, i.e., 
         \begin{eqnarray}
            \label{Security II}
            &\text{[Security II]}& \ \ \ I(\mathcal{D}, \mathcal{W};\ X^{\text{mirror}}_{1},\ldots,X^{\text{mirror}}_{K_1})=0.
        \end{eqnarray}
        Besides, the following colluding attack from users is considered. For any two subsets
        $\mathcal{T}_1\subseteq [K_1]$ and $\mathcal{T}_2\subseteq [K_2]$, each user of $\mathcal{U}_{\mathcal{T}_1,\mathcal{T}_2}$ can not obtain any information about the request vectors of other users in $\mathcal{U}_{K_1]\setminus\mathcal{T}_1,[K_2]\setminus\mathcal{T}_2}$, i.e., 
        \begin{eqnarray}
            \label{Privacy II}
            \text{[Privacy II]} \ I(\mathcal{D}_{[K_1]\backslash\mathcal{T}_1,[K_2]\backslash\mathcal{T}_2}; \{X^{\text{mirror}}_{k_1}\}_{k_1\in\mathcal{T}_1},\mathcal{D}_{\mathcal{T}_1,\mathcal{T}_2}, \widetilde{\mathcal{Z}}_{\mathcal{T}_1,\mathcal{T}_2},\mathcal{W})=0, \forall \mathcal{T}_1\subseteq [K_1],\forall \mathcal{T}_2\subseteq [K_2].
        \end{eqnarray}
    \end{itemize}


 Finally, in a hierarchical caching system, each user $\text{U}_{k_1,k_2}$ is able to retrieve its demand linear function $L_{\mathbf{d}_{k_1,k_2}}$. Then the following equation must hold.
\begin{eqnarray}
\label{Decodability}
&\text{[Decodability]}& \ \ \  H(L_{\mathbf{d}_{k_1,k_2}}|\ \ X^{\text{mirror}}_{k_1},\ \mathbf{d}_{k_1,k_2},\ \widetilde{\mathcal{Z}}_{k_1,k_2}) = 0,  \ \ \ \forall k_1\in[K_1],\  \forall k_2\in[K_2].
\end{eqnarray}



In this paper, we will consider the cases of with and without privacy and security constraints, respectively. Our goal is to find novel coded caching schemes to reduce the transmission loads ($R_1,R_2$) or subpacketization for the worst case, i.e., the demanding matrix $\mathbf{D}$ is full-rank, and establish the optimality results for some scenarios. Besides, we intend to find how to modify the scheme obtained from HPDA into a secure and private scheme (SP-scheme), i.e., satisfies the constraints \eqref{Security I},\eqref{Privacy I},\eqref{Security II},\eqref{Privacy II}.

\section{Hierarchical Placement Delivery Array}
\label{sec:HPDA}
In this section, we first briefly describe the vanilla PDA for the single-layer cache-aided broadcast network \cite{YCTC}, and then introduce a new concept named HPDA, which can be used to characterize the placement and delivery strategies of coded caching schemes with or without security and privacy for the hierarchical caching system in Fig. \ref{fig-model}.
\subsection{Placement Delivery Array}
\label{subsec:PDA}
\begin{definition}
\label{def-PDA}
(\cite{YCTC}) For any positive integers $K,F, Z$ and $S$, an $F\times K$ array $\mathbf{P}=(p_{j,k})_{j\in[F] ,k\in[K]}$ over alphabet set $\{*\}\bigcup [S]$ is called a $(K,F,Z,S)$ PDA if it satisfies the following conditions.
 \item [C$1$.] The symbol ``$*$" appears $Z$ times in each column;
 \item [C$2$.] Each integer occurs at least once in the array;
 \item [C$3$.] For any two distinct entries $p_{j_1,k_1}$ and $p_{j_2,k_2}$, $p_{j_1,k_1}=p_{j_2,k_2}=s$ is an integer only if
 \begin{enumerate}
 \item [a.] $j_1\ne j_2$, $k_1\ne k_2$, i.e., they lie in distinct rows and distinct columns; and
  \item [b.] $p_{j_1,k_2}=p_{j_2,k_1}=*$, i.e., the corresponding $2\times 2$ subarray formed by rows $j_1,j_2$ and columns $k_1,k_2$ must be of the following form
 \begin{align*}
 \left(\begin{array}{cc}
 s & *\\
 * & s
 \end{array}\right)~\textrm{or}~
 \left(\begin{array}{cc}
 * & s\\
 s & *
 \end{array}\right).
 \end{align*}
 \end{enumerate}
\end{definition}
\begin{example}
\label{example-1}
When $K=4$, $F=6$, $Z=3$ and $S=4$, we can see that the following array is a $(4,6,3,4)$ PDA.
\begin{eqnarray}
\label{eg-PDA()3313}
\mathbf{A}=\left(\small{
     \begin{array}{cccc}
       \ast & \ast & 1 & 2 \\
       \ast & 1 & \ast & 3 \\
       \ast & 2 & 3 &\ast \\
       1 & \ast & \ast & 4 \\
       2 & \ast & 4 & \ast \\
       3 & 4 & \ast & \ast \\
     \end{array}}
   \right).
   \end{eqnarray}
\end{example}
Given a $(K,F,Z,S)$ PDA, the row labels represent the file packet indexes and the column labels represent the user indexes. If the entry $p_{j,k}=*$, it means user $k$ has cached the $j$-th packet of all files. Condition C$1$ of Definition \ref{def-PDA} implies that all users have the same memory ratio of $\frac{M}{N}=\frac{Z}{F}$. If the entry $p_{j,k}=s$, the $j$-th packet of all files is not cached by user $k$. Then server transmit a coded signal which is generated by the XOR of all the requested packets indicated by $s$ to users. Condition C$2$ of Definition \ref{def-PDA} implies that the number of signals sent by the server is exactly $S$. Condition C$3$ of Definition \ref{def-PDA} guarantees that each user can decode its requested packets, since it has cached all the packets in the signal except for its requested one. The authors in \cite{YCTC} showed that a $(K,F,Z,S)$ PDA can be used to realize an $F$-division $(K,M,N)$ coded caching scheme with $\frac{M}{N}=\frac{Z}{F}$ and transmission load $R=\frac{S}{F}$ for the single-layer cache-aided broadcast network.
Furthermore the seminal coded cahcing scheme proposed in~\cite{MN} can be represented by a special PDA which is referred to as MN PDA. That is the following result.
\begin{lemma}\rm(MN PDA\cite{MN,YCTC})
\label{le-MN}
For any positive integers $K$ and $t$ with $t\leq K$, there exists a   $\left(K,{K\choose t},{K-1\choose t-1},{K\choose t+1}\right)$ PDA which realizes a $(K,M,N)$ MN scheme with $\frac{M}{N}=\frac{t}{K}$, subpacketization $F={K\choose t}$ and transmission load $R=\frac{K-t}{t+1}$.
\hfill $\square$
\end{lemma}
Here we briefly review the construction of MN PDA as follows.
\begin{construction}\rm (MN PDA\cite{MN,YCTC})
\label{con-MN} For any integer $t\in[K]$, let $F={K\choose t}$. Then we have a ${K\choose t}\times K$ array $\mathbf{P}=\left(\mathbf{P}(\mathcal{T},k)\right)_{\mathcal{T}\in {[K]\choose t}, k\in [K]}$ by
\begin{align}\label{Eqn_Def_AN}
\mathbf{P}(\mathcal{T},k)=\left\{\begin{array}{cc}
\phi_{t+1}(\mathcal{T}\cup\{k\}), & \mbox{if}~k\notin\mathcal{T}\\
*, & \mbox{otherwise},
\end{array}
\right.
\end{align}
where $\phi_{t+1}(\cdot)$ is a bijection from  $\binom{[K]}{t+1}$ to $[\binom{K}{t+1}]$ and the rows are labelled by all the subsets $\mathcal{T}\in {[K]\choose t}$ listed in the lexicographic order. For instance, the PDA in \eqref{eg-PDA()3313} is a $(4,6,3,4)$ MN PDA.
\hfill $\square$
\end{construction}

Finally we should point out that PDA has been widely studied. There are some schemes with lower subpacketization level based on PDA proposed in \cite{YCTC,CJYT,CJWY,CWZW,WCWG,MJW,ZCJ,ZCW,CWLZG,SSRS}. In addition, the authors in \cite{SKTADA} pointed out that all the proposed schemes in \cite{TR,SZG,STD,YTCC,KP} could be represented by appropriate PDAs.

\subsection{Hierarchical Placement Delivery Array and Examples}
Inspired by PDA, for the hierarchical cache-aided network, we propose another combination structure, namely HPDA, which can be used to characterize the placement and delivery strategies of a coded caching scheme. The definition of hierarchical placement delivery array (HPDA) is given as follows.
\label{subsec:HPDA}
\begin{definition}\label{def-H-PDA}
For any positive integers $K_{1}, K_{2}, F$, $Z_{1}$, $Z_{2}$ with $Z_1<F$, $Z_2<F$ and any integer sets $\mathcal{S}_{\text{M}}$ and $\mathcal{S}_{k_1}$, $k_1\in[K_1]$, an $F\times (K_1+K_1K_2)$ array $\mathbf{A}=(\mathbf{A}^{(0)},\mathbf{A}^{(1)},\ldots,\mathbf{A}^{(K_1)})$,
where $\mathbf{A}^{(0)}=(a^{(0)}_{j,k_1})_{j\in[F],k_1\in [K_1]}$ is an $F\times K_1$ array consisting of $*$ and null, and $\mathbf{A}^{(k_1)}=(a^{(k_1)}_{j,k_2})_{j\in[F],k_2\in [K_2]}$ is an $F\times K_2$ array over $\{*\}\bigcup \mathcal{S}_{k_1}$,  $k_1\in[K_1]$, is a $(K_1,K_2;F;Z_1,Z_2;\mathcal{S}_{\text{M}}$, $\mathcal{S}_1,\ldots,\mathcal{S}_{K_1})$ hierarchical placement delivery array (HPDA) if it satisfies the following conditions.
\begin{itemize}
\item[B1.] Each column of $\mathbf{A}_0$ has $Z_1$ stars;
\item[B2.] $\mathbf{A}^{(k_1)}$ is a $(K_2,F,Z_2,|\mathcal{S}_{k_1}|)$ PDA for each $k_1\in [K_1]$;
\item[B3.] Each integer $s\in \mathcal{S}_{\text{M}}$ occurs in exactly one subarray $\mathbf{A}^{(k_1)}$, and for each $a^{(k_1)}_{j,k_2}=s\in \mathcal{S}_{\text{M}}$, we have  $a^{(0)}_{j,k_1}=*$ where $k_1\in[K_1]$, and $j\in[F]$, $k_1\in[K_1]$, $k_2\in[K_2]$;
\item[B4.] For any two entries $a^{(k_1)}_{j,k_2}$ and $a^{(k'_1)}_{j',k'_2}$ where $k_1\neq k'_1\in[K_1]$, $j,j'\in [F]$ and $k_2,k'_2\in[K_2]$, if $a^{(k_1)}_{j,k_2}=a^{(k'_1)}_{j',k'_2}=s$, then
\begin{itemize}
\item $a^{(k_1)}_{j',k_2}$ is an integer only if $a^{(0)}_{j',k_1}=*$;
\item $a^{(k'_1)}_{j,k'_2}$ is an iteger only if $a^{(0)}_{j,k'_1}=*$.
\end{itemize}
\end{itemize}
\end{definition}

Similar to PDA, given an $F\times(K_1+K_1K_2)$ HPDA $\mathbf{A}=(\mathbf{A}^{(0)},\mathbf{A}^{(1)},\ldots,\mathbf{A}^{(K_1)})$, the row labels also represent the packet indexes; the column labels of $\mathbf{A}^{(0)}$ and $\mathbf{A}^{(k_1)}$, $k_1\in[K_1]$ represent the mirror site indexes and user indexes respectively, and the stars also have the same representation as the star in PDA. The integers also represent the required coded packets. However there are two types of integers in $\mathbf{A}$. For the integers $s\in\mathcal{S}_{\text{M}}$, they indicate the requested packets only sent by mirror sites, as illustrated in Condition B$3$ of Definition \ref{def-H-PDA}. For the integers $s\in\bigcup^{K_1}_{k_1=1}\mathcal{S}_{k_1}\setminus\mathcal{S}_{\text{M}}$, they indicate the requested packets sent from the server, and Condition B$4$ of Definition \ref{def-H-PDA} guarantees that each user can decode their requested packet with the help of the packets cached by its connected mirror site or itself.
From a $(K_1,K_2;F;Z_1,Z_2;\mathcal{S}_{\text{M}},\mathcal{S}_1,\ldots,\mathcal{S}_{K_1})$ HPDA, we can also obtain an $F$-$(K_1,K_2; M_1,M_2; N)$ hierarchical coded caching scheme. First we take the following example to illustrate the coded caching scheme obtained from an HPDA.

\begin{example}[The scheme without security and privacy constraints]
\label{ex-2}
When $K_{1}=2$, $K_{2}=2$, $F=6$, $Z_{1}=1$, $Z_{2}=2$, $\mathcal{S}_{\text{M}}=[5:8], \mathcal{S}_1=[6], \mathcal{S}_2=[4]\bigcup[7:8]$,  one can check that the following array is a $(2,2;6;1,2;\mathcal{S}_{\text{M}},\mathcal{S}_1,\mathcal{S}_2)$ HPDA $\mathbf{A}$ in \eqref{eg-HPDA}.
\begin{eqnarray}
\label{eg-HPDA}
\begin{split}
\mathbf{A}&=(\mathbf{A}_0, \mathbf{A}_1, \mathbf{A}_2)\\
&=
\left(
\begin{array}{cc|cc|cc}
  * &   & 5 & 6 & 1 & 2   \cr
    &   & * & 1 & * & 3   \cr
    &   & * & 2 & 3 & *   \cr
    &   & 1 & * & * & 4   \cr
    &   & 2 & * & 4 & *   \cr
    & * & 3 & 4 & 7 & 8
\end{array}
\right).
\end{split}
\end{eqnarray}
We can get an $F$-$(K_1,K_2;M_1,M_2;N)=4$-$(2,2;4,8;24)$ coded caching scheme based on \eqref{eg-HPDA} in the following way.
\begin{itemize}
	\item \textbf{Placement Phase}: Each file from $\mathbb{F}^{B}_q$ is divided into $F=6$ packets with equal size, i.e., $W_{n}=\{W_{n,1},W_{n,2},\ldots,W_{n,6}\}, n\in [24]$. Because $a^{(0)}_{1,1}=a^{(0)}_{6,2}=*$ in \eqref{eg-HPDA}, the contents cached by mirror sites are as follows:
\begin{eqnarray}
\label{cache-mir-ori}
&&\mathcal{Z}_{1} =\{W_{n,1}\ |\ n\in[24]\},\ \mathcal{Z}_{2} =\{W_{n,6}\ |\ n\in[24]\}.
\end{eqnarray}
Because $a^{(1)}_{2,1}=a^{(1)}_{3,1}=a^{(1)}_{4,2}=a^{(1)}_{5,2}=a^{(2)}_{2,1}=a^{(2)}_{4,1}=a^{(2)}_{3,2}=a^{(2)}_{5,2}=*$ in \eqref{eg-HPDA}, the packets cached by the users are as follows:
\begin{eqnarray}
\label{cache-user-ori}
&&\widetilde{\mathcal{Z}}_{1,1} =\{W_{n,2}, W_{n,3}\ |\ n\in[24]\}, \
\widetilde{\mathcal{Z}}_{1,2} =\{W_{n,4}, W_{n,5}\ |\ n\in[24]\},\nonumber \\
&&\widetilde{\mathcal{Z}}_{2,1} =\{W_{n,2}, W_{n,4}\ |\ n\in[24]\}, \
\widetilde{\mathcal{Z}}_{2,2} =\{W_{n,3}, W_{n,5}\ |\ n\in[24]\}.
\end{eqnarray}
\item\textbf{Delivery Phase}: Assume that $\text{U}_{1,1}$ requests $W_1+W_2$, $\text{U}_{1,2}$ requests $W_3+W_4$, $\text{U}_{2,1}$ requests $W_5+W_6$ and $\text{U}_{2,2}$ requests $W_7+W_8$, then the demand matrix is
\begin{eqnarray}
\label{eq-request-matrix}
\mathbf{D}=\left(\small{
     \begin{array}{c}
       \mathbf{D}_{1} \\
       \hline
       \mathbf{D}_{2}
     \end{array}}
   \right)=\left(\small{
     \begin{array}{c}
       \mathbf{d}_{1,1} \\
       \mathbf{d}_{1,2} \\
       \hline
       \mathbf{d}_{2,1} \\
       \mathbf{d}_{2,2}
     \end{array}}
   \right)=\left(\small{
     \begin{array}{c}
       1,1,0,0,0,0,0,0,\ldots,0 \\
       0,0,1,1,0,0,0,0,\ldots,0 \\
       \hline
       0,0,0,0,1,1,0,0,\ldots,0 \\
       0,0,0,0,0,0,1,1,\ldots,0
     \end{array}}
   \right)_{4\times24}.
\end{eqnarray}
For any vector $\mathbf{v}=(v_1,\ldots,v_{24})$, we use the following notation to denote a linear combination of packets.
\begin{eqnarray}
\label{def-request-linear-packet}
    L_{\mathbf{v},j} &:=& \sum\limits^{24}\limits_{n=1}v_n\cdot W_{n,j},j\in[6].
\end{eqnarray}

The messages sent to all users consist of two types.
\begin{itemize}
\item {\bf{The messages $X^{\text{server}}_{s}$ sent by server:}} For each $s\in(\mathcal{S}_1\bigcup\mathcal{S}_2)\setminus\mathcal{S}_\text{M}=[4]$, server sends a coded signal $X^{\text{server}}_{s}$. Similar to the transmission of the scheme realized by PDA, the server sends the following coded packets.
\begin{eqnarray}
\label{eq-packet-server}
\begin{split}
&X^{\text{server}}_{1}=L_{\mathbf{d}_{1,1},4}+L_{\mathbf{d}_{1,2},2}+L_{\mathbf{d}_{2,1},1},\ \ \ \
X^{\text{server}}_{2}=L_{\mathbf{d}_{1,1},5}\!+\!L_{\mathbf{d}_{1,2},3}\!+\!L_{\mathbf{d}_{2,2},1},&\\
&X^{\text{server}}_{3}=L_{\mathbf{d}_{1,1},6}\!+\!L_{\mathbf{d}_{2,1},3}\!+\!L_{\mathbf{d}_{2,2},2},\ \ \ \ \ \  X^{\text{server}}_{4}=L_{\mathbf{d}_{1,2},6}\!+\!L_{\mathbf{d}_{2,1},5}\!+\!L_{\mathbf{d}_{2,2},4}.&
\end{split}
\end{eqnarray} Server transmits these $4$ coded packets to all mirror sites. So the transmission load of the first layer is $R_1=\frac{4}{6}=2/3$.
\item {\bf The messages $X^{\text{mirror}}_{k_1,s}$ sent by $\text{M}_{k_1}$:} We take $\text{M}_{1}$ as an example. For each $s\in\mathcal{S}_{1}\setminus\mathcal{S}_{\text{M}}=[4]$, then mirror site $\text{M}_{1}$ sends a processed coded signal $X^{\text{mirror}}_{1,s}$ by canceling all the packets in $X^{\text{server}}_{s}$ which have been cached by $\text{M}_{1}$ and not requested by user in $\mathcal{U}_{1}$. For instance the mirror site $\text{M}_{1}$ sends $X^{\text{mirror}}_{1,1}$ by subtracting $L_{\mathbf{d}_{2,1},1}$ in $X^{\text{server}}_{1}$ which is cached in $\mathcal{Z}_1$ and not requested by user $\mathbf{U}_{1,1}$ or $\mathbf{U}_{1,2}$. That is,
  \begin{eqnarray}
        \label{eq-packet-mirror}
        X^{\text{mirror}}_{1,1}=X^{\text{server}}_{1}-L_{\mathbf{d}_{2,1},1} =L_{\mathbf{d}_{1,1},4}+L_{\mathbf{d}_{1,2},2}.
    \end{eqnarray}
  Similarly $\text{M}_{1}$ sends the coded signal $X^{\text{mirror}}_{1,2}$ by canceling $L_{\mathbf{d}_{2,2},1}$ in $X^{\text{server}}_{2}$ and directly sends $X^{\text{mirror}}_{1,3}$, $X^{\text{mirror}}_{1,4}$ as follows.
        \begin{eqnarray*}
        X^{\text{mirror}}_{1,2}&=&X^{\text{server}}_{2}\!-\!L_{\mathbf{d}_{2,2},1}=L_{\mathbf{d}_{1,1},5}+L_{\mathbf{d}_{1,2},3},\\
        X^{\text{mirror}}_{1,3}&=&X^{\text{server}}_{3},\ \  X^{\text{mirror}}_{1,4}=X^{\text{server}}_{4}.
        \end{eqnarray*}

For each $s'\in \mathcal{S}_{1}\bigcap \mathcal{S}_{\text{M}}=[5:6]$, the mirror site $\text{M}_{1}$ directly generates coded signals $X^{\text{mirror}}_{1,s'}$ from $\mathcal{Z}_{1}$ by means of the delivery strategy realized by PDA $\mathbf{A}_{1}$. $\text{M}_{1}$ sends $X^{\text{mirror}}_{1,5}$, $X^{\text{mirror}}_{1,6}$ to users $\text{U}_{1,1}$ and $\text{U}_{1,2}$ as follows.
        \begin{eqnarray}
        \label{eq-packet2-mirror}
            X^{\text{mirror}}_{1,5}=L_{\mathbf{d}_{1,1},1}\ \
            X^{\text{mirror}}_{1,6}=L_{\mathbf{d}_{1,2},1}
        \end{eqnarray}
We can check that user $\text{U}_{1,1}$ can decode $W_{1,4}+W_{2,4}$, $W_{1,5}+W_{2,5}$, $W_{1,6}+W_{2,6}$, $W_{1,1}+W_{2,1}$ respectively from the received coded packets $X^{\text{mirror}}_{1,1},X^{\text{mirror}}_{1,2},X^{\text{mirror}}_{1,3},X^{\text{mirror}}_{1,5}$ since it has cached packets $\{W_{n,2}$, $W_{n,3} |\ n\in[24]\}$. Similarly each user is able to recover its desired linear combination of files, and the transmission amount by mirror site is $6$ packets, so the transmission load of the second layer is $R_2=\frac{6}{6}=1$.
\end{itemize}
\end{itemize}
\end{example}
In Example \ref{ex-2}, we show that from an HPDA we can obtain a coded caching scheme for a hierarchical caching system which only achieves \eqref{Decodability}, i.e., the scheme without security and prvacy. In fact, by adding some random variables into the placement phase and using some other random variables to encode the transmitted signals in the delivery phase, we can also obtain a scheme satisfying \eqref{Security I}-\eqref{Decodability}, i.e., the SP-scheme in Example \ref{ex-3}.
\begin{example}[The scheme with security and privacy constraints]
\label{ex-3}
Based on the same HPDA in \eqref{eg-HPDA}, we can get an $F$-$(K_1,K_2;M_1$, $M_2;N)=4$-$(2,2;13/3,26/3;24)$ SP-scheme by modifying Example \ref{ex-2} in the following way.

\begin{itemize}
	\item \textbf{Placement Phase}: Each file is still divided into $6$ packets. In order to guarantee the security constraints in \eqref{Security I}, \eqref{Security II} and privacy constraints in \eqref{Privacy I}, \eqref{Privacy II}, the variable $P$ consists of two classes of variables which are independent to each other. The first class contains some security vectors. That is, the server generates $|\mathcal{S}_1\cup \mathcal{S}_2|=8$ security vectors uniformly and independently from $\mathbb{F}^{B/6}_q$, i.e., $\{\mathbf{V}_s |\ \mathbf{V}_s\in\mathbb{F}^{B/6}_q, s\in[8]\}$. The second class contains some privacy vectors. That is, the server generates $K_1K_2=4$ privacy vectors $ \mathbf{p}_{1,1},\mathbf{p}_{1,2},\mathbf{p}_{2,1},\mathbf{p}_{2,2}$ for each user uniformly and independently at random from $\mathbb{F}^{24}_q$ as follows.
\begin{eqnarray*}
\mathbf{p}_{1,1}\!=\!(p^{(1)}_{1,1},p^{(2)}_{1,1},\ldots,p^{(24)}_{1,1})\
\mathbf{p}_{1,2}\!=\!(p^{(1)}_{1,2},p^{(2)}_{1,2},\ldots,p^{(24)}_{1,2})\\
\mathbf{p}_{2,1}\!=\!(p^{(1)}_{2,1},p^{(2)}_{2,1},\ldots,p^{(24)}_{2,1})\
\mathbf{p}_{2,2}\!=\!(p^{(1)}_{2,2},p^{(2)}_{2,2},\ldots,p^{(24)}_{2,2}).
\end{eqnarray*} Besides of the contents cached by the mirror sites in \eqref{cache-mir-ori}, each mirror site $\text{M}_{k_1}$ additionally caches security vector $\mathbf{V}_s$ if $s$ occurs in $\mathcal{S}_{k_1}\bigcap\mathcal{S}_{\text{M}}$ for all $s\in [8]$. That is, $\text{M}_{1}$ caches $\mathbf{V}_5,\mathbf{V}_6$, and $\text{M}_{2}$ caches $\mathbf{V}_7,\mathbf{V}_8$, because $\mathcal{S}_{1}\bigcap\mathcal{S}_{\text{M}}=[5:6]$, and $\mathcal{S}_{2}\bigcap\mathcal{S}_{\text{M}}=[7:8]$. Then the cache contents in \eqref{cache-mir-ori} can be written as follows.
\begin{eqnarray*}
\mathcal{Z}_{1} =\{W_{n,1}\ |\ n\in[24]\}\bigcup\{\mathbf{V}_5, \mathbf{V}_6\} \ \ \ \
\mathcal{Z}_{2} =\{W_{n,6}\ |\ n\in[24]\}\bigcup\{\mathbf{V}_7, \mathbf{V}_8\}.
\end{eqnarray*}
In order to guarantee that each user can decode its desired linear combination packets from the encrypted signal, each user additionally caches some coded contents. We consider user $\text{U}_{1,1}$ as an example. Clearly there are four integers in the first column of $\mathbf{A}_1$, i.e., $a^{(1)}_{1,1}=5$, $a^{(1)}_{4,1}=1$, $a^{(1)}_{5,1}=2$, $a^{(1)}_{6,1}=3$. Then using security vectors $\mathbf{V}_s$ where $s\in\{1,2,3,5\}$ and privacy vector $\mathbf{p}_{1,1}$, user $\text{U}_{1,1}$ caches additional coded packets
$\{\mathbf{V}_5+L_{\mathbf{p}_{1,1},1}$,$\mathbf{V}_1+L_{\mathbf{p}_{1,1},4}$,$ \mathbf{V}_2+L_{\mathbf{p}_{1,1},5}$,$\mathbf{V}_3+L_{\mathbf{p}_{1,1},6}\}$.
Similarly the other users also caches some coded packets respectively, and \eqref{cache-user-ori} can be written as follows.
\begin{eqnarray*}
&&\widetilde{\mathcal{Z}}_{1,1} =\{W_{n,2}, W_{n,3}\ |\ n\in[24]\}\bigcup\left\{\mathbf{V}_5\!+\!L_{\mathbf{p}_{1,1},1},\mathbf{V}_1\!+\!L_{\mathbf{p}_{1,1},4}, \mathbf{V}_2\!+\!L_{\mathbf{p}_{1,1},5},\mathbf{V}_3\!+\!L_{\mathbf{p}_{1,1},6}\right\},\\
&&\widetilde{\mathcal{Z}}_{1,2} =\{W_{n,4}, W_{n,5}\ |\ n\in[24]\}\bigcup\left\{\mathbf{V}_6\!+\!L_{\mathbf{p}_{1,2},1}, \mathbf{V}_1\!+\!L_{\mathbf{p}_{1,2},2}, \mathbf{V}_2\!+\!L_{\mathbf{p}_{1,2},3}, \mathbf{V}_4\!+\!L_{\mathbf{p}_{1,2},6}\right\},\\
&&\widetilde{\mathcal{Z}}_{2,1} =\{W_{n,2}, W_{n,4}\ |\ n\in[24]\}\bigcup\left\{\mathbf{V}_1\!+\!L_{\mathbf{p}_{2,1},1}, \mathbf{V}_3\!+\!L_{\mathbf{p}_{2,1},3}, \mathbf{V}_4\!+\!L_{\mathbf{p}_{2,1},5}, \mathbf{V}_7\!+\!L_{\mathbf{p}_{2,1},6}\right\},\\
&&\widetilde{\mathcal{Z}}_{2,2} =\{W_{n,3}, W_{n,5}\ |\ n\in[24]\}\bigcup\left\{\mathbf{V}_2\!+\!L_{\mathbf{p}_{2,2},1}, \mathbf{V}_3\!+\!L_{\mathbf{p}_{2,2},2}, \mathbf{V}_4\!+\!L_{\mathbf{p}_{2,2},4}, \mathbf{V}_8\!+\!L_{\mathbf{p}_{2,2},6}\right\}.
\end{eqnarray*}
\item\textbf{Delivery Phase}:
We also consider the demand matrix in \eqref{eq-request-matrix}. Then using the privacy vectors and request vectors, the server generates the following public vector set $\mathcal{Q}=\{ \mathbf{q}_{1,1},\mathbf{q}_{1,2},\mathbf{q}_{2,1},\mathbf{p}_{2,2}\}$ where
\begin{eqnarray}
\label{eq-public-vector}
\mathbf{q}_{1,1}=\mathbf{p}_{1,1}+\mathbf{d}_{1,1}=\left(p^{(1)}_{1,1}+1, p^{(2)}_{1,1}+1,p^{(3)}_{1,1}, p^{(4)}_{1,1},p^{(5)}_{1,1},p^{(6)}_{1,1},p^{(7)}_{1,1},p^{(8)}_{1,1}\ldots,p^{(24)}_{1,1}\right),\nonumber\\
\mathbf{q}_{1,2}=\mathbf{p}_{1,2}+\mathbf{d}_{1,2}=\left(p^{(1)}_{1,2}, p^{(2)}_{1,2}, p^{(3)}_{1,2}+1, p^{(4)}_{1,2}+1,p^{(5)}_{1,2},p^{(6)}_{1,2},p^{(7)}_{1,2},p^{(8)}_{1,2}\ldots,p^{(24)}_{1,2}\right),\nonumber\\
\mathbf{q}_{2,1}=\mathbf{p}_{2,1}+\mathbf{d}_{2,1}=\left(p^{(1)}_{2,1}, p^{(2)}_{2,1},p^{(3)}_{2,1}, p^{(4)}_{2,1},p^{(5)}_{2,1}+1,p^{(6)}_{2,1}+1,p^{(7)}_{2,1},p^{(8)}_{2,1}\ldots,p^{(24)}_{2,1}\right),\nonumber\\
\mathbf{q}_{2,2}=\mathbf{p}_{2,2}+\mathbf{d}_{2,2}=\left(p^{(1)}_{2,2}, p^{(2)}_{2,2}, p^{(3)}_{2,2}, p^{(4)}_{2,2},p^{(5)}_{2,2},p^{(6)}_{2,2},p^{(7)}_{2,2}+1,p^{(8)}_{2,2}+1\ldots,p^{(24)}_{2,2}\right).
\end{eqnarray}	
Note that $\mathbf{q}_{k_1,k_2}$ are independent from each other due to the independent random vectors $\mathbf{p}_{k_1,k_2}$, $k_1\in[K_1]$, $k_2\in[K_2]$.
The messages sent to all users also consist of two types.
\begin{itemize}
\item {\bf{The messages $X^{\text{server}}_{s}$ sent by the server:}} For each $s\in(\mathcal{S}_1\bigcup\mathcal{S}_2)\setminus\mathcal{S}_\text{M}=[4]$, the server sends a coded signal $X^{\text{server}}_{s}$ consisting of the the coded packets in \eqref{eq-packet-server} together with the secure vectors and some privacy contents. For instance, the server sends the signal $X^{\text{server}}_{1}$ as follows.
\begin{align}
X^{\text{server}}_{1}&=\underbrace{\mathbf{V}_1}_{\text{secure vector}}+\underbrace{L_{\mathbf{p}_{1,1},4}+L_{\mathbf{p}_{1,2},2}+L_{\mathbf{p}_{2,1},1}}_{\text{privacy contents}}+\underbrace{L_{\mathbf{d}_{1,1},4}+L_{\mathbf{d}_{1,2},2}+L_{\mathbf{d}_{2,1},1}}_{\text{the coded signal $X^{\text{server}}_{1}$ in \eqref{eq-packet-server}}}\nonumber\\
&=\mathbf{V}_1+(L_{\mathbf{p}_{1,1},4}+L_{\mathbf{d}_{1,1},4})+(L_{\mathbf{p}_{1,2},2}+L_{\mathbf{d}_{1,2},2})+(L_{\mathbf{p}_{2,1},1}+L_{\mathbf{d}_{2,1},1})\nonumber\\
&\stackrel{(a)}= ~\mathbf{V}_1\!+\!L_{\mathbf{q}_{1,1},4}\!+\! L_{\mathbf{q}_{1,2},2}\!+\!L_{\mathbf{q}_{2,1},1}.\label{eq-packet-sp-server}
\end{align}
The equation (a) holds because by \eqref{def-request-linear-packet} and \eqref{eq-public-vector} we have $L_{\mathbf{p}_{k_1,k_2},j}+L_{\mathbf{d}_{k_1,k_2},j}=L_{\mathbf{q}_{k_1,k_2},j}$, where $k_1\in[2]$, $k_2\in[2]$, $j\in[6]$. Similarly, the rest signals are written as follows.
\begin{eqnarray*}
X^{\text{server}}_{2}&=&\mathbf{V}_2\!+\!L_{\mathbf{q}_{1,1},5}\!+\! L_{\mathbf{q}_{1,2},3}\!+\!L_{\mathbf{q}_{2,2},1},\
X^{\text{server}}_{3}=\mathbf{V}_3\!+\!L_{\mathbf{q}_{1,1},6}\!+\! L_{\mathbf{q}_{2,1},3}\!+\!L_{\mathbf{q}_{2,2},2}\\
X^{\text{server}}_{4}&=&\mathbf{V}_4\!+\!L_{\mathbf{q}_{1,2},6}\!+\! L_{\mathbf{q}_{2,1},5}\!+\!L_{\mathbf{q}_{2,2},4}
\end{eqnarray*}	Server transmits the coded messages $\mathcal{Q}$,  $X^{\text{server}}_{1}$, $X^{\text{server}}_{2}$, $X^{\text{server}}_{3}$, $X^{\text{server}}_{4}$ to all mirror sites. Compared with the length of file packets, the length of public vectors in $\mathcal{Q}$ is short enough to neglect, so the transmission load of the first layer is $R_1=\frac{4}{6}=2/3$.
\item {\bf{The messages $X^{\text{mirror}}_{k_1,s}$ sent by $\text{M}_{k_1}$:}} We take $\text{M}_{1}$ as an example. For each $s\in \mathcal{S}_{1}\setminus\mathcal{S}_{\text{M}}=[4]$, the mirror site $\text{M}_{1}$ sends a processed coded signal $X^{\text{mirror}}_{1,s}$ by canceling all the packets in $X^{\text{server}}_{s}$ which have been cached by $\text{M}_{1}$ and not requested by any user in $\mathcal{U}_{1}$. For instance $\text{M}_{1}$ sends $X^{\text{mirror}}_{1,1}$ by subtracting $L_{\mathbf{q}_{2,1},1}$ in $X^{\text{server}}_{1}$ which is cached in $\mathcal{Z}_1$ and not requested by user $\mathbf{U}_{1,1}$ or $\mathbf{U}_{1,2}$. Similar to \eqref{eq-packet-mirror}, we have
    \begin{eqnarray}
    \label{eq-packet-sp-mirror}
    X^{\text{mirror}}_{1,1}=X^{\text{server}}_{1}\!-\!L_{\mathbf{q}_{2,1},1}
    =\mathbf{V}_1\!+\!(L_{\mathbf{d}_{1,1},4}\!+\!L_{\mathbf{p}_{1,1},4})\!+\!(L_{\mathbf{d}_{1,2},2}\!+\!L_{\mathbf{p}_{1,2},2}).
    \end{eqnarray}
In the same manner, $\text{M}_{1}$ transmits signal $X^{\text{mirror}}_{1,2}$ by canceling $L_{\mathbf{q}_{2,2},1}$ in $X^{\text{server}}_{2}$ and directly sends $X^{\text{mirror}}_{1,3}$ and $X^{\text{mirror}}_{1,4}$ to users in $\mathcal{U}_{1}$ as follows.
\begin{eqnarray*}
X^{\text{mirror}}_{1,2}&=&X^{\text{server}}_{2}-L_{\mathbf{q}_{2,2},1}
=\mathbf{V}_2+\left(L_{\mathbf{d}_{1,1},5}+L_{\mathbf{p}_{1,1},5}\right)+\left(L_{\mathbf{d}_{1,2},3}+ L_{\mathbf{p}_{1,2},3}\right),\\
X^{\text{mirror}}_{1,3}&=&X^{\text{server}}_{3}=\mathbf{V}_3+(L_{\mathbf{d}_{1,1},6}+L_{\mathbf{p}_{1,1},6})+(L_{\mathbf{d}_{2,1},3}+L_{\mathbf{p}_{2,1},3})+(L_{\mathbf{d}_{2,2},2}+L_{\mathbf{p}_{2,2},2}),\\
X^{\text{mirror}}_{1,4}&=&X^{\text{server}}_{4}=\mathbf{V}_4+(L_{\mathbf{d}_{1,2},6}+L_{\mathbf{p}_{1,2},6})+(L_{\mathbf{d}_{2,1},5}+L_{\mathbf{p}_{2,1},5})+(L_{\mathbf{d}_{2,2},4}+L_{\mathbf{p}_{2,2},4}).
\end{eqnarray*}

 For each $s'\in \mathcal{S}_{1}\bigcap\mathcal{S}_{\text{M}}=[5:6]$, mirror site $\text{M}_{1}$ directly generates coded signals $X^{\text{mirror}}_{1,s'}$ by means of sending the coded packets in \eqref{eq-packet2-mirror} together with some secure vectors and some privacy contents, then we have
    \begin{eqnarray*}
    X^{\text{mirror}}_{1,5}=\underbrace{\mathbf{V}_5}_{\text{secure vector}}+\underbrace{L_{\mathbf{p}_{1,1},1}}_{\text{privacy contens}}+\underbrace{L_{\mathbf{d}_{1,1},1}}_{\text{the coded signal $X^{\text{mirror}}_{1,5}$ in \eqref{eq-packet2-mirror} }}=\mathbf{V}_5+L_{\mathbf{q}_{1,1},1}.
    \end{eqnarray*}
    Similarly, the rest signal is $ X^{\text{mirror}}_{1,6}=\mathbf{V}_6+L_{\mathbf{p}_{1,2},1}+L_{\mathbf{d}_{1,2},1}=\mathbf{V}_6+L_{\mathbf{q}_{1,2},1}$.

    We can check that user $\text{U}_{1,1}$ can  decode $W_{1,4}+W_{2,4}$, $W_{1,5}+W_{2,5}$, $W_{1,6}+W_{2,6}$, $W_{1,1}+W_{2,1}$ from $X^{\text{mirror}}_{1,1}$, $X^{\text{mirror}}_{1,2}$, $X^{\text{mirror}}_{1,3}$, $X^{\text{mirror}}_{1,5}$, because it has cached packets $\{W_{n,2}$, $W_{n,3} |\ n\in[24]\}\bigcup\left\{\mathbf{V}_5\!+\!L_{\mathbf{p}_{1,1},1},\mathbf{V}_1\!+\!L_{\mathbf{p}_{1,1},4}, \mathbf{V}_2\!+\!L_{\mathbf{p}_{1,1},5},\mathbf{V}_3\!+\!L_{\mathbf{p}_{1,1},6}\right\}$ and it knows $\mathcal{Q}$ (see detailed analysis in Section \ref{scheme-decodability}). Similarly, each user is able to recover its desired linear combination of files, and the transmission amount by mirror site is still $6$ packets, so the transmission load of the second layer is $R_2=\frac{6}{6}=1$.

\end{itemize}
\end{itemize}
Because each transmitted coded packet is encrypted by the security vector $\mathbf{V}_s$ with the same length of the file packet, in terms of the information theory, the security constraints \eqref{Security I}, \eqref{Security II} can be guaranteed. Furthermore, since the public vectors  $\mathbf{q}_{1,1},\mathbf{q}_{1,2},\mathbf{q}_{2,1},\mathbf{q}_{2,2}$ are independently and uniformly distributed over $\mathbb{F}^{24}_{q}$, then privacy constraints \eqref{Privacy I}, \eqref{Privacy II} are satisfied. We will propose the formal proof of the security and privacy in Section \ref{sec:proof-th-1}.
\end{example}
By the above two examples, they have the same transmission loads of $R_1$, $R_2$ and subpacketization while the memory ratios of the scheme in Example \ref{ex-2}, i.e., $\frac{M_1}{N}=\frac{12}{72}$, $\frac{M_2}{N}=\frac{24}{72}$,  would be a little lower than the SP-scheme ($\frac{M_1}{N}=\frac{13}{72}$, $\frac{M_2}{N}=\frac{26}{72}$). That is because both the two schemes has the same memory of uncoded content (the file packets labeled by $*$), but SP-scheme has to cache some extra coded contents, which only works for protecting the security and privacy of the delivery. Based on an HPDA, the following results can be obtained.

\begin{theorem}
\label{th-1}
Given a $(K_1,K_2;F;Z_1,Z_2;\mathcal{S}_{\text{M}},\mathcal{S}_1,\ldots,\mathcal{S}_{K_1}) $ HPDA $\mathbf{A}=(\mathbf{A}_0,\mathbf{A}_1,\ldots,\mathbf{A}_{K_1})$,  we can obtain an $F$-division $(K_1,K_2;M_1,M_2;N)$ coded caching scheme with security and privacy for a linear function retrieval problem with $\frac{M_1}{N}=\frac{Z_1}{F}+\frac{|\mathcal{S}_{\text{M}}\bigcap\mathcal{S}_{k_1}|}{NF}$, $\frac{M_2}{N}=\frac{Z_2}{F}+\frac{F-Z_2}{NF}$ and transmission load $R_1=\frac{|\bigcup_{k_1=1}^{K_1}\mathcal{S}_{k_1}|-|\mathcal{S}_{\text{M}}|}{F}$, $R_2=\max_{k_1\in[K_1]}\left\{\  \frac{\mid\mathcal{S}_{k_1}\mid}{F}\ \right\}$. \qed
\end{theorem}

The detailed proof is included in Section \ref{sec:proof-th-1}. According to the proof\footnote{In Section \ref{sec:proof-th-1} when we delete the security vectors and privacy vectors, the cache packets for the mirror sites and users are obtained in Remark \ref{remark-4}, and transmitted coded packets are obtained in Remark \ref{remark-5}. By Remark \ref{remark-6}, we can see each user can decode its required linear combination packets.} of Theorem \ref{th-1}, we can also obtain a scheme without the security and privacy constraints \eqref{Security I}-\eqref{Privacy II} as follows.

\begin{corollary}
\label{coro-th-1}
Given a $(K_1,K_2;F;Z_1,Z_2;\mathcal{S}_{\text{M}},\mathcal{S}_1,\ldots,\mathcal{S}_{K_1}) $ HPDA $\mathbf{A}=(\mathbf{A}_0,\mathbf{A}_1,\ldots,\mathbf{A}_{K_1})$,  we can obtain an $F$-division $(K_1,K_2;M_1,M_2;N)$ coded caching scheme without security and privacy for a linear function retrieval problem with $\frac{M_1}{N}=\frac{Z_1}{F}$, $\frac{M_2}{N}=\frac{Z_2}{F}$ and transmission load $R_1=\frac{|\bigcup_{k_1=1}^{K_1}\mathcal{S}_{k_1}|-|\mathcal{S}_{\text{M}}|}{F}$, $R_2=\max_{k_1\in[K_1]}\left\{\  \frac{\mid\mathcal{S}_{k_1}\mid}{F}\ \right\}$. \qed
\end{corollary}
\begin{remark}
When the file number $N$ is large, the cost of extra caching memory for the security and privacy can be neglect, i.e., $\frac{Z_1}{F}+\frac{|\mathcal{S}_{\text{M}}\bigcap\mathcal{S}_{k_1}|}{NF}\approx\frac{Z_1}{F}$ and $\frac{Z_2}{F}+\frac{F-Z_2}{NF}\approx\frac{Z_2}{F}$.\qed
\end{remark}

\section{Main Results and Performance Analysis}
\label{sec:main-result}

From Definition \ref{def-PDA} and Definition \ref{def-H-PDA}, we can use PDA to generate HPDA. In this section we first propose two classes of constructions of HPDA. Consequently we obtain two classes of HPDA which lead to two classes of coded caching scheme with or without security and privacy for hierarchical network. Finally the theoretical and numerical performance analyses are proposed.

\subsection{The Construction of Grouping Method}
\label{cons-th2}
\begin{figure}[http!]
\centering
\includegraphics[scale=0.9]{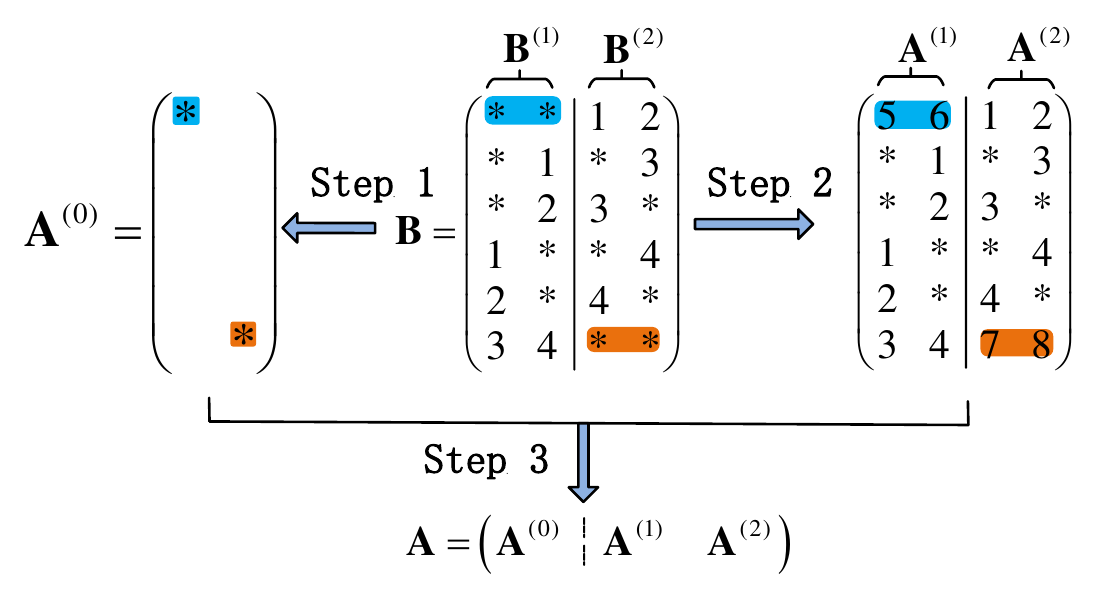}
\caption{ The sketch of transformation from MN PDA $\mathbf{B}$ to a HPDA $\mathbf{A}$ in Theorem \ref{th-2}.}
\label{fig-sketch2}
\end{figure}
In this subsection, we introduce the idea of constructing HPDA $\mathbf{A}= \left(\mathbf{A}^{(0)}\right.$, $\mathbf{A}^{(1)}$, $\ldots$, $\left.\mathbf{A}^{(K_1)}\right)$ by  grouping method based on a PDA. First the following notation is useful. Given any array,  we call a row of it \emph{star row} if this row contains only star entries.  The construction of grouping method can be briefly described as follows: Given a PDA $\mathbf{B}=\left(\mathbf{B}^{(1)},\ldots,\mathbf{B}^{(K_1)}\right)$, to construct $F\times K_1$ array $\mathbf{A}^{(0)}=(a^{(0)}_{f,k_1})_{f\in[F], k_1\in[K_1]}$, we let  its element  $a^{(0)}_{f,k_1}$    be a star entry   if the $f$-th row of $\mathbf{B}^{(k_1)}$ is a star row, and be   null otherwise. To construct the array   $\mathbf{A}^{(k_1)}$, we simply replace all the star entries in each star row of   $\mathbf{B}^{(k_1)}$ with distinct integers,  for all $k_1\in[K_1]$. Note that if a PDA satisfies the condition that it can be divided into several groups which have the same number of star row, the PDA is suitable for the grouping method. We take the following example to show the construction of $\mathbf{A}$ based on a MN PDA.

Given a $(K_1K_2,$$F,Z,S)$$=$$(2\times2,$$6,3,4)$ MN PDA $\mathbf{B}=\left(\mathbf{B}^{(1)},\ldots,\mathbf{B}^{(2)}\right)$, i.e., the PDA in \eqref{eg-PDA()3313}, we will use a grouping method to construct a $(2,2;6;1,2;$ $\mathcal{S}_{\text{M}},$$ \mathcal{S}_1,$$ \mathcal{S}_2)$ HPDA $\mathbf{A}$ in \eqref{eg-HPDA}, where
\begin{eqnarray}
\label{eq-alphabet-set2}
&&\mathcal{S}_{\text{M}}=[5:8],\
\mathcal{S}_\text{1}=[1:6],\
\mathcal{S}_\text{2}=[1:4]\cup [7:8].
\end{eqnarray}
 The construction includes the following three steps, as illustrated in Fig. \ref{fig-sketch2}.
\begin{itemize}
\item\textbf{Step 1.} Construction of  $\mathbf{A}^{(0)}$$=(a^{(0)}_{f,k_1})_{f\in[6], k_1\in[2]}$ for mirror sites. Since there is exactly one star row in $\mathbf{B}^{(1)}$, i.e., the first row, we fill $a^{(0)}_{1,1}=*$, and for the rest entries in column $1$ of $\mathbf{A}^{(0)}$, we fill them with null. Similarly, the column $2$ of $\mathbf{A}^{(0)}$ can be obtained by applying the same operation on $\mathbf{B}^{(2)}$. Then $\mathbf{A}^{(0)}$ is obtained.
\item\textbf{Step 2.} Construction of $\left(\mathbf{A}^{(1)},\mathbf{A}^{(2)}\right)=$ $(a^{(k_1)}_{f,k_2})$ for users, where $f\in[6]$, $k_2\in[2]$, $k_1\in[2]$. Taking $\mathbf{A}^{(1)}$ as an example, in the star row of $\mathbf{B}^{(1)}$, let $b^{(1)}_{1,1}=5$, $b^{(1)}_{1,2}=6$ to obtain $\mathbf{A}^{(1)}$. Similarly we can obtain $\mathbf{A}^{(2)}$ in Fig. \ref{fig-sketch2}. So as shown in  Fig. \ref{fig-sketch2} we have the integer sets $\mathcal{S}_1$ and $\mathcal{S}_2$ of $\mathbf{A}^{(1)}$ and $\mathbf{A}^{(2)}$ in \eqref{eq-alphabet-set2}.

\item\textbf{Step 3.} Construction of $\mathbf{A}$. We get a $6\times 6$ array by arranging $\mathbf{A}^{(0)}$ and $\mathbf{A}^{(1)}$ horizontally, i.e. $\mathbf{A}=\left(\mathbf{A}^{(0)}, \mathbf{A}^{(1)}, \mathbf{A}^{(2)}\right)$.
\end{itemize}

Now we show that the obtained $\mathbf{A}$ satisfies the conditions of Definition \ref{def-H-PDA}. It is easy to check that B1 and B2 of Definition \ref{def-H-PDA} hold. So we only need to check B3 and B4. From Fig. \ref{fig-sketch2}, we have $\mathcal{S}_{\text{M}}=[5:8]$, whose integers only appear in one $\mathbf{A}^{(k_1)},k_1\in[2]$, and we can check that, if $a^{(k_1)}_{f,k_2}=s\in\mathcal{S}_{\text{M}}$, then $a^{(0)}_{f,k_1}=*$, thus   Condition B$3$ of Definition \ref{def-H-PDA} holds. It can be checked that Condition B$4$ also holds. Take $a^{(k_1)}_{f,k_2}=a^{(k'_1)}_{f',k'_2}=1$ as an example, where $k_1\neq k'_1$.  From \eqref{eg-HPDA}, we can see that $a^{(1)}_{4,1}=a^{(1)}_{2,2}=a^{(2)}_{1,1}=1$. When choosing $f=4,k_1=k_2=1$ and $f'=k'_2=1$, $k'_1=2$, we have $a^{(k_1)}_{f',k_2}=5$, and the corresponding $a^{(0)}_{f',k_1}$ equals to $*$, i.e., $a^{(0)}_{1,1}=*$, satisfying Condition B$4$ of Definition \ref{def-H-PDA}.

From the above construction, the required HPDA $\mathbf{A}$ which is the array in \eqref{eg-HPDA} is obtained. By Example \ref{ex-2}, we have a $4$-$(2,2;4,8;24)$ coded caching scheme with $R_1=\frac{2}{3}$. By Example \ref{ex-3}, we have a $4$-$(2,2;13/3,26/3;24)$ SP-scheme with $R^{\text{sp}}_1=R_1=\frac{2}{3}$. In a $(2,2;4,8;24)$ hierarchical caching system without security and privacy, by the exhaustive computer searches for the values of $\alpha$ and $\beta$ which result in the optimal $R_1$ in \eqref{DHCC-B}, \eqref{eq-load-W}, we have the minimum loads $R^{\text{KNMD}}_1=0.687$ from \eqref{DHCC-B} and $R^{\text{WWCY}}_1=0.738$  from \eqref{eq-load-W} respectively in \cite{KNMD,WWCY}. Clearly $R_1<R^{\text{KNMD}}_1<R^{\text{WWCY}}_1$.

Generally, for the grouping method of constructing HPDA based on a MN PDA, we have the following results.
\begin{theorem}
\label{th-2}
    For any positive integers $K_1$, $K_2$, $t$, $K_2\leq t\leq K_1K_2$, there exists   a $(K_1,K_2$; ${K_1K_2\choose t}$; ${K_1K_2-K_2\choose t-K_2}$, ${K_1K_2-1\choose t-1}-{K_1K_2-K_2\choose t-K_2}$; $\mathcal{S}_{\text{M}},\mathcal{S}_1$, $\ldots$, $\mathcal{S}_{K_1})$ HPDA,  which leads to an $F$-division $(K_1$, $K_2$; $M_1$, $M_2$; $N)$ SP-scheme for a linear function retrieval problem with
\begin{subequations}
\label{th-2-para}
\begin{IEEEeqnarray}{rCl}
      &&\text{Memory ratios}:  \frac{M_1}{N}=\frac{{K_1K_2-K_2\choose t-K_2}}{{K_1K_2\choose t}}+\frac{K_2{K_1K_2-K_2\choose t-K_2}}{N{K_1K_2\choose t}} ,\label{eqRatioThm2_1} \\
      && \ \ \ \ \ \ \ \ \ \ \ \ \ \ \ \ \ \ \ \frac{M_2}{N}=\frac{t}{K_1K_2}-\frac{{K_1K_2-K_2\choose t-K_2}}{{K_1K_2\choose t}}+\frac{{K_1K_2\choose t}-{K_1K_2-1\choose t-1}+{K_1K_2-K_2\choose t-K_2}}{N{K_1K_2\choose t}},~\quad\label{eqRatioThm2_2} \\[0.2cm]
      &&\text{Subpacketization}:  F={K_1K_2\choose t},\label{eqPackThm2}\\[0.2cm]
      &&\text{Transmission
    loads}:  R_1=\frac{{K_1K_2-t}}{{t+1}},\label{eqR1Thm2} R_2=\frac{{K_1K_2-t}}{{t+1}}-\frac{{K_1K_2-K_2\choose t+1}}{{K_1K_2\choose t}}+\frac{{K_1K_2-K_2\choose t-K_2}K_2}{{K_1K_2\choose t}}.\label{eqR2Thm2}
\end{IEEEeqnarray}
\end{subequations}\qed
\end{theorem}
By Corollary \ref{coro-th-1} and the HPDA in Theorem \ref{th-2}, we have the following result.
\begin{corollary}
\label{coro-th-2}
 For any positive integers $K_1$, $K_2$, $t$, $K_2\leq t\leq K_1K_2$, there exists   a $(K_1,K_2$; ${K_1K_2\choose t}$; ${K_1K_2-K_2\choose t-K_2}$, ${K_1K_2-1\choose t-1}-{K_1K_2-K_2\choose t-K_2}$; $\mathcal{S}_{\text{M}},\mathcal{S}_1$, $\ldots$, $\mathcal{S}_{K_1})$ HPDA,  which leads to an $F$-division $(K_1$, $K_2$; $M_1$, $M_2$; $N)$ coded caching scheme without security and privacy for a linear function retrieval problem with
\begin{subequations}
\begin{IEEEeqnarray}{rCl}
  \label{th-para2}
      &&\text{Memory ratios}:  \frac{M_1}{N}=\frac{{K_1K_2-K_2\choose t-K_2}}{{K_1K_2\choose t}},~\label{eqRatioThm22 1}  \frac{M_2}{N}=\frac{t}{K_1K_2}-\frac{{K_1K_2-K_2\choose t-K_2}}{{K_1K_2\choose t}},~\quad\label{eqRatioThm22 2} \\[0.2cm]
      &&\text{Subpacketization}:  F={K_1K_2\choose t},\label{eqPackThm22}\\[0.2cm]
      &&\text{Transmission
    loads}:  R_1=\frac{{K_1K_2-t}}{{t+1}},\label{eqR1Thm22} R_2=\frac{{K_1K_2-t}}{{t+1}}-\frac{{K_1K_2-K_2\choose t+1}}{{K_1K_2\choose t}}+\frac{{K_1K_2-K_2\choose t-K_2}K_2}{{K_1K_2\choose t}}.\label{eqR2Thm22}
\end{IEEEeqnarray}
\end{subequations}
\qed
\end{corollary}

Note that single file retrieval is included in linear function retrieval problem. Then the optimal worst-case load under the constraint of uncoded data placement is not decreased when the users request scalar linear functions of the files and the demand matrix is row full rank. By deriving the optimal worst-case load under uncoded placement and the assumption that each user requests one single file, we can obtain the following result whose proof is included in Appendix \ref{appendix-optimal}.

\begin{lemma}
\label{Lemma-optimality}
The scheme for a two-level hybrid network under the uncoded data placement in Corollary \ref{coro-th-2} has the \emph{optimal load} $R_1$ for the first layer, when the demand matrix is row full rank.
\qed
\end{lemma}

\begin{remark}
\label{remark-2}
From the sketch of construction of the HPDA in Fig. \ref{fig-sketch2}, the proposed HPDA generated from grouping method can be regarded as assigning the stars in each column of a MN PDA to mirror sites and users. It is natural that the first layer load $R_1$ in \eqref{eqR1Thm22} equals to the load of the scheme realized by the MN PDA. Note that this star assigning strategy guarantees the sufficient utilization of stars in decoding the multicast signal sent by the server, which minimize the transmission load $R_1$. However, it also leads to a increase in $R_2$, because for the desired packets cached by mirror sites, they are sent to each user by the unicast transmission.
\end{remark}

As one can see, not every PDA can be used to construct HPDA by the grouping method, which results in a parameter constraint. In order to have a more flexible choice of the parameters, we put forward another construction called hybrid method in the next subsection.

\subsection{The Construction of Hybrid Method}
\begin{figure}[http!]
\centering
\includegraphics[scale=0.9]{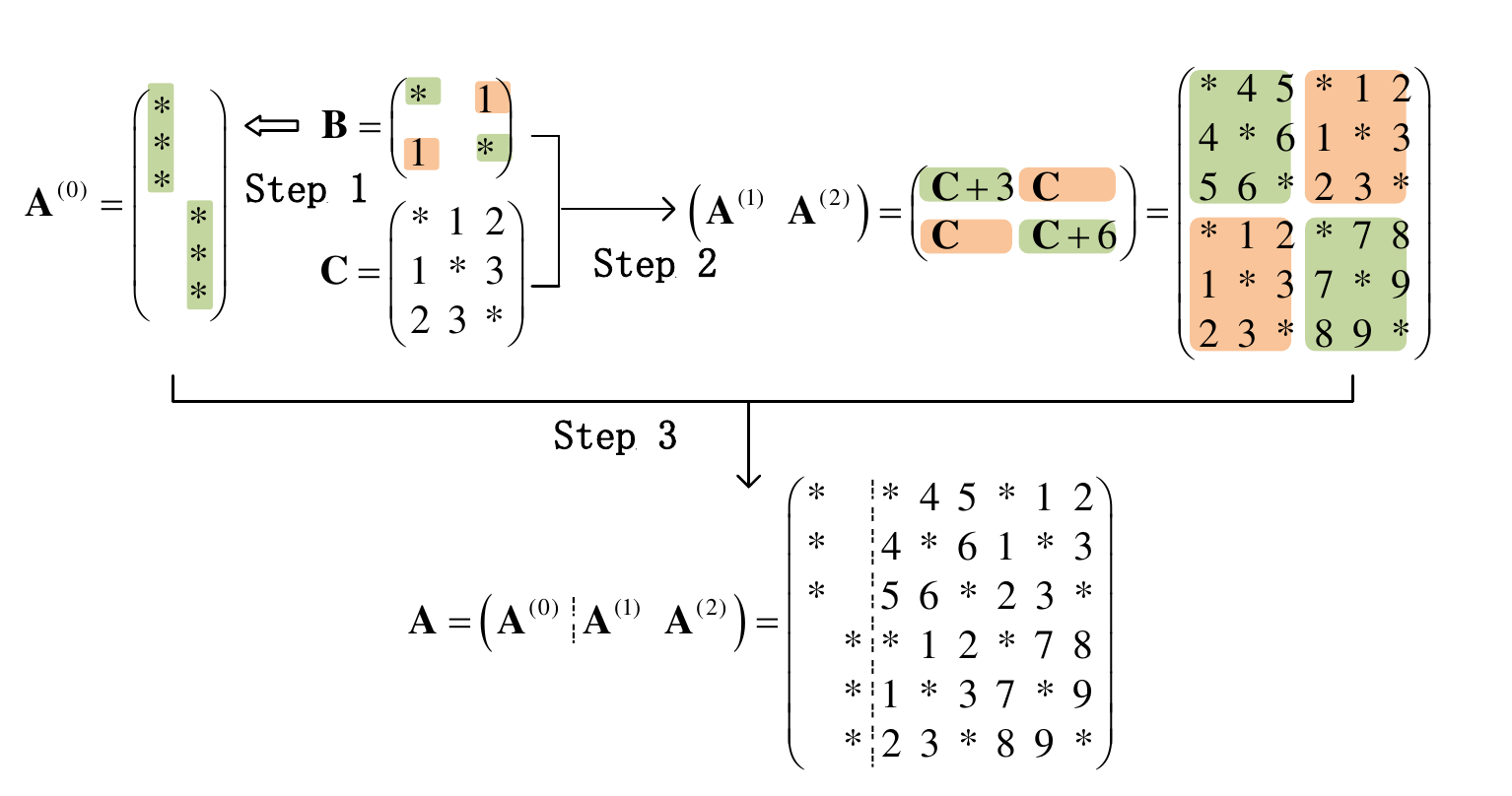}
\caption{ The sketch of ptransformation from MN PDAs $\mathbf{B}$ and $\mathbf{C}$ to a HPDA $\mathbf{A}$ in Theorem \ref{th-3}.}
\label{fig-sketch}
\end{figure}
In this subsection, we introduce the idea of constructing HPDA $\mathbf{A}= \left(\mathbf{A}^{(0)}\right.$, $\mathbf{A}^{(1)}$, $\ldots$, $\left.\mathbf{A}^{(K_1)}\right)$ by a hybrid method based on two PDAs. The idea of hybrid method can be briefly described as follows: Given two PDAs, denoted as $\mathbf{B}$, $\mathbf{C}$ respectively, $\mathbf{A}^{(0)}$ is obtained by deleting all the integers of $\mathbf{B}$ and then simply expanding each row of it, and
$\left(\mathbf{A}^{(1)}\right.$, $\ldots$, $\left.\mathbf{A}^{(K_1)}\right)$ is obtained by using a hybrid method, in which $\mathbf{B}$ acts as an outer array and the inner arrays are simply constructed by $\mathbf{C}$. We take the following example to show the construction of $\mathbf{A}$ based on two MN PDA.

Given a $(K_1,F_1,Z_1,S_1)=(2,2,1,1)$ MN PDA $\mathbf{B}=(b_{f_1,k_1})_{f_1\in[2], k_1\in[2]}$ and a $(K_2,F_2,Z_2,S_2)=(3,3,1,3)$ MN PDA $\mathbf{C}=(c_{f_2,k_2})_{f_2\in[3], k_2\in[3]}$ as follows.
\begin{eqnarray}
\label{eq-two-array}
\mathbf{B}=\left(
\begin{array}{cc}
* & 1 \\
1 & *
\end{array}
\right) \ \  \ \
\mathbf{C}=\left(
\begin{array}{ccc}
*	&	1	&	2\\
1	&	*	&	3\\
2	&	3	&	*
\end{array}
\right).
\end{eqnarray}
 We will use a hybrid method to construct a $(2,3;6;3,2;$ $\mathcal{S}_{\text{M}}$, $\mathcal{S}_\text{1},\mathcal{S}_\text{2})$ HPDA where
\begin{eqnarray}
\label{eq-alphabet-set}
\mathcal{S}_{\text{M}}=[4:9],\ \
\mathcal{S}_\text{1}=[1:6],\ \
\mathcal{S}_\text{2}=[1:3]\cup [7:9],
\end{eqnarray}
 through the following three steps, as illustrated in Fig. \ref{fig-sketch}.
\begin{itemize}
\item\textbf{Step 1.} Construction of $\mathbf{A}^{(0)}$ for mirror sites. We can get a $6\times 2$ array $\mathbf{A}^{(0)}$ by deleting all the integer entries of $\mathbf{B}=(b_{f_1,k_1})_{f_1\in[2], k_1\in[2]}$ and then expanding each row $3$ times\footnote{\label{foot;ICM(20,2)}
		This is the row number of the inner structure array which will be introduced in Step 2.}.
\item\textbf{Step 2.} Construction of $\left(\mathbf{A}^{(1)},\mathbf{A}^{(2)}\right)$ for users. We replace the integer entries $b_{2,1}$ and $b_{1,2}$ by $\mathbf{C}$ ($\mathbf{C}$ is of the size of inner structure array ), and replace $b_{1,1}=b_{2,2}=*$ by $\mathbf{C}+3$, $\mathbf{C}+6$ respectively to get
\begin{eqnarray}
\label{eq-subarrays}
    \begin{split}
        \left(
        \begin{array}{c|c}
            \mathbf{A}^{(1)}&\mathbf{A}^{(2)}
        \end{array}
        \right)
        =\left(
        \begin{array}{c|c}
            \mathbf{C}+3&\mathbf{C}\\
            \mathbf{C}&\mathbf{C}+6
        \end{array}
        \right).
    \end{split}
\end{eqnarray}
\item\textbf{Step 3.} Construction of $\mathbf{A}$. We get a $6\times 8$ array by arranging $\mathbf{A}^{(0)}$ and $(\mathbf{A}^{(1)},\mathbf{A}^{(2)})$ horizontally, i.e., $\mathbf{A}=\left(\mathbf{A}^{(0)},\mathbf{A}^{(1)},\mathbf{A}^{(2)}\right)$.
\end{itemize}
Now we verify that the construction above leads to an HPDA defined in Definition \ref{def-H-PDA}. It's easy to check that Conditions B$1$ and B$2$ of Definition \ref{def-H-PDA} hold. Then we only need to check Conditions B$3$ and B$4$. From Fig. \ref{fig-sketch}, we have $\mathcal{S}_{\text{M}}=[4:9]$, whose integers only appear in one $\mathbf{A}^{(k_1)}$, $k_1\in[2]$, and we can check that, if $a^{(k_1)}_{f,k_2}=s\in\mathcal{S}_{\text{M}}$, then $a^{(0)}_{f,k_1}=*$, $k_1\in[2]$, $k_2\in[3]$, $f\in[6]$, thus Condition B$3$ of Definition \ref{def-H-PDA} holds.
For Condition B$4$ of Definition \ref{def-H-PDA}, here we take $a^{(k_1)}_{f,k_2}$$=a^{(k'_1)}_{f',k'_2}$$=1$ as an example, where $k_1\neq k'_1$. From Fig. \ref{fig-sketch}, we can see that $a^{(1)}_{5,1}$ $=a^{(1)}_{4,2}$ $=a^{(2)}_{2,1}$ $=a^{(2)}_{1,2}=1$. Choosing $f=5$, $k_1=k_2=1$ and $f'=k'_1=2$, $k'_2=1$, we have $a^{(k_1)}_{f',k_2}=4$, and the corresponding $a^{(0)}_{f',k_1}$ equals to $*$, i.e., $a^{(0)}_{2,1}=*$, satisfying Condition B$4$ of Definition \ref{def-H-PDA}.

Generally, for the hybrid method of constructing HPDA based on two PDAs, we have the following results.
\begin{theorem}
\label{th-3}
For any $(K_1,F_1,Z_1,S_1)$ PDA $\mathbf{B}$ and $(K_2,F_2,Z_2,S_2)$ PDA $\mathbf{C}$, there exists a $(K_1,K_2$; $F_1F_2$; $Z_1F_2$, $Z_2F_1$; $\mathcal{S}_{\text{M}}$, $\mathcal{S}_1$, $\ldots$, $\mathcal{S}_{K_1})$ HPDA $\mathbf{A}$, which leads to an $F_1F_2$-division $(K_1,K_2;M_1,M_2;N)$ SP-scheme for a linear function retrieval problem where:
\begin{subequations}
\label{th-3-para}
\begin{IEEEeqnarray}{rCl}
      &&\text{Memory ratios}:  \frac{M_1}{N}=\frac{Z_1}{F_1}+\frac{Z_1S_2}{NF_1F_2} ,\label{eqRatioThm3_1}   \frac{M_2}{N}=\frac{Z_2}{F_2}+\frac{F_2-Z_2}{F_2N},~\quad\label{eqRatioThm3_2} \\[0.2cm]
      &&\text{Subpacketization}:  F=F_1F_2,\label{eqPackThm3}\\[0.2cm]
      &&\text{Transmission
    loads}:  R_1=\frac{S_1S_2}{F_1F_2}, R_2=\frac{S_2}{F_2}.\label{eqR2Thm3}
\end{IEEEeqnarray}
\end{subequations}\qed
\end{theorem}

By Corollary \ref{coro-th-1} and the HPDA in Theorem \ref{th-3}, we have the following result.
\begin{corollary}
\label{coro-th-3}
For any $(K_1,F_1,Z_1,S_1)$ PDA $\mathbf{B}$ and $(K_2,F_2,Z_2,S_2)$ PDA $\mathbf{C}$, there exists a $(K_1,K_2$; $F_1F_2$; $Z_1F_2$, $Z_2F_1$; $\mathcal{S}_{\text{M}}$, $\mathcal{S}_1$, $\ldots$, $\mathcal{S}_{K_1})$ HPDA $\mathbf{A}$, which leads to an $F_1F_2$-division $(K_1,K_2;M_1,M_2;N)$ coded caching scheme without security and privacy for a linear function retrieval problem where:
\begin{subequations}
\begin{IEEEeqnarray}{rCl}
  \label{th3-para}
      &&\text{Memory ratios}:  \frac{M_1}{N}=\frac{Z_1}{F_1},\label{eqRatioThm32_1}~\frac{M_2}{N}=\frac{Z_2}{F_2},~\quad\label{eqRatioThm32_2} \\[0.2cm]
      &&\text{Subpacketization}:  F=F_1F_2,\label{eqPackThm32}\\[0.2cm]
      &&\text{Transmission
    loads}:  R_1=\frac{S_1S_2}{F_1F_2},~R_2=\frac{S_2}{F_2}.\label{eqR2Thm32}
\end{IEEEeqnarray}
\end{subequations}\qed
\end{corollary}

\begin{remark}\label{remark-3}
Unlike the strong limitations on parameters and a high subpacketization of the schemes in Theorem \ref{th-2} and Corollary \ref{coro-th-2},  the schemes in Theorem \ref{th-3} and Corollary \ref{coro-th-3} have a more flexible choices and lower subpacketizations. By choosing different PDAs $\mathbf{B}$ and $\mathbf{C}$ to construct the HPDA by using a hybrid construction, we can obtain schemes with different transmission loads and subpacketization. In particular, the following statements hold.

\begin{itemize}
\item
When $\mathbf{B}$ and $\mathbf{C}$ both are MN PDAs, and the hierarchical caching system is without security and privacy and each user requests one single file, the corresponding scheme is the same as the WWCY scheme \cite{WWCY}, which achieves the optimal transmission load of $R^*_2$;
\item When $\mathbf{B}$ is the PDA proposed in \cite{YCTC} and $\mathbf{C}$ is a MN PDA, the corresponding scheme could reduce the subpacketization at the cost of increasing communication loads. We name it as Scheme I for Theorem \ref{coro-th-3};
\item When  both $\mathbf{B}$ and $\mathbf{C}$  are   PDAs proposed in \cite{YCTC},  the corresponding scheme can further reduce the  subpacketization. We name it as Scheme II for Theorem \ref{coro-th-3}.
\end{itemize}\qed
\end{remark}

\begin{figure}[http!]
\centering
\subfigure[]
{
\includegraphics[width=5cm]{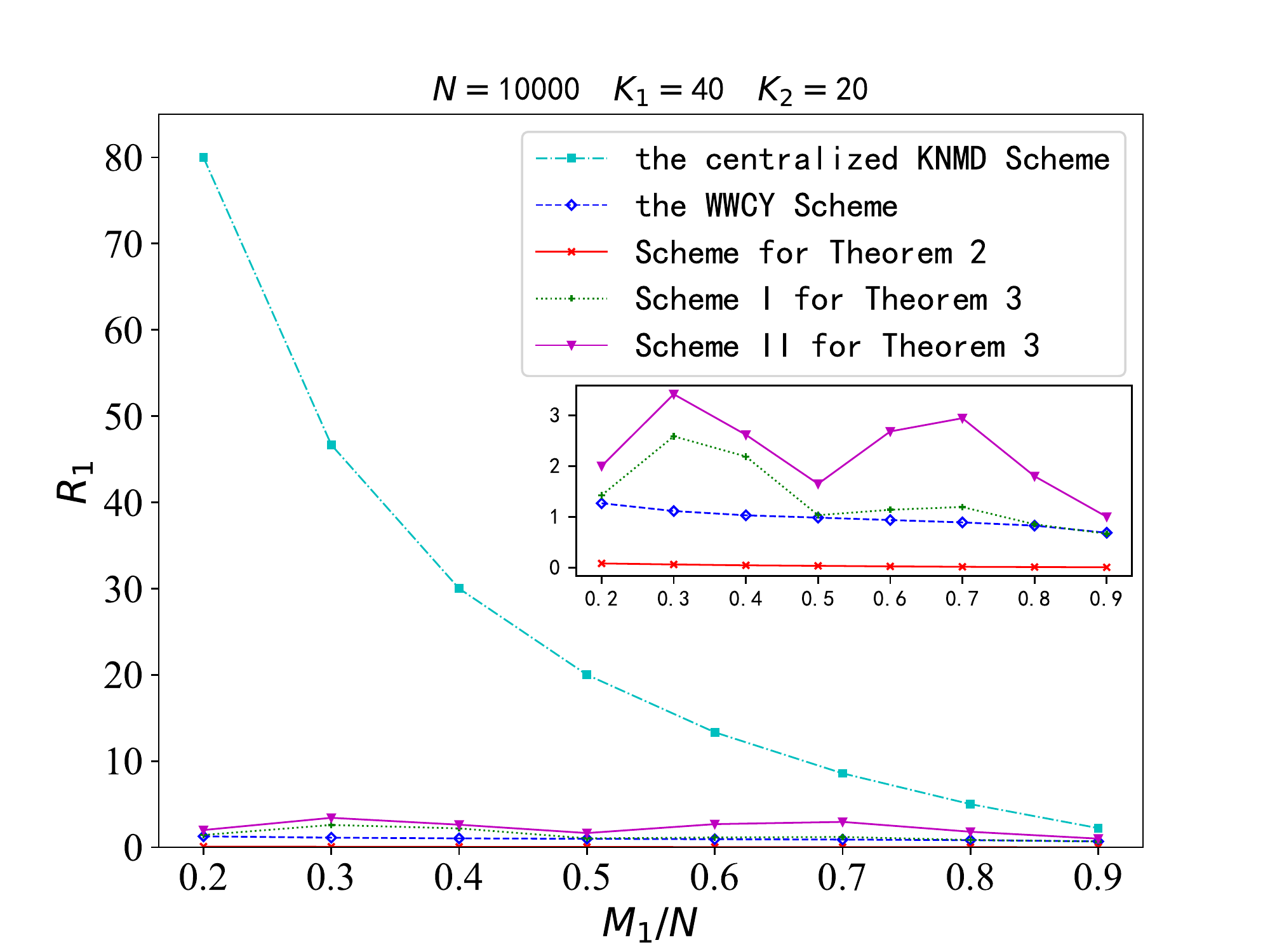}
\label{subfig1}
}
\subfigure[]{
\includegraphics[width=5cm]{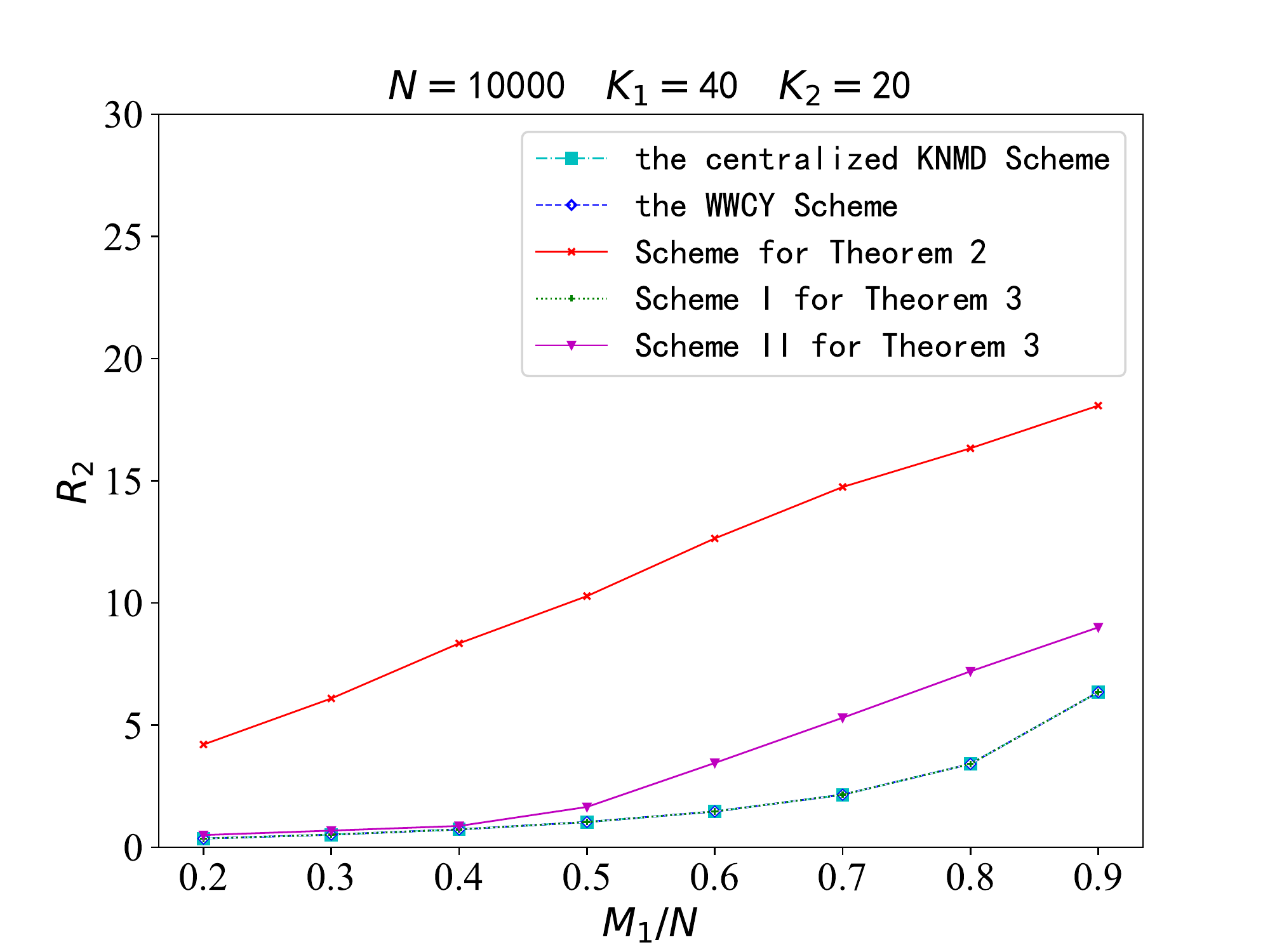}
\label{subfig2}
}
\subfigure[]{
\includegraphics[width=5cm]{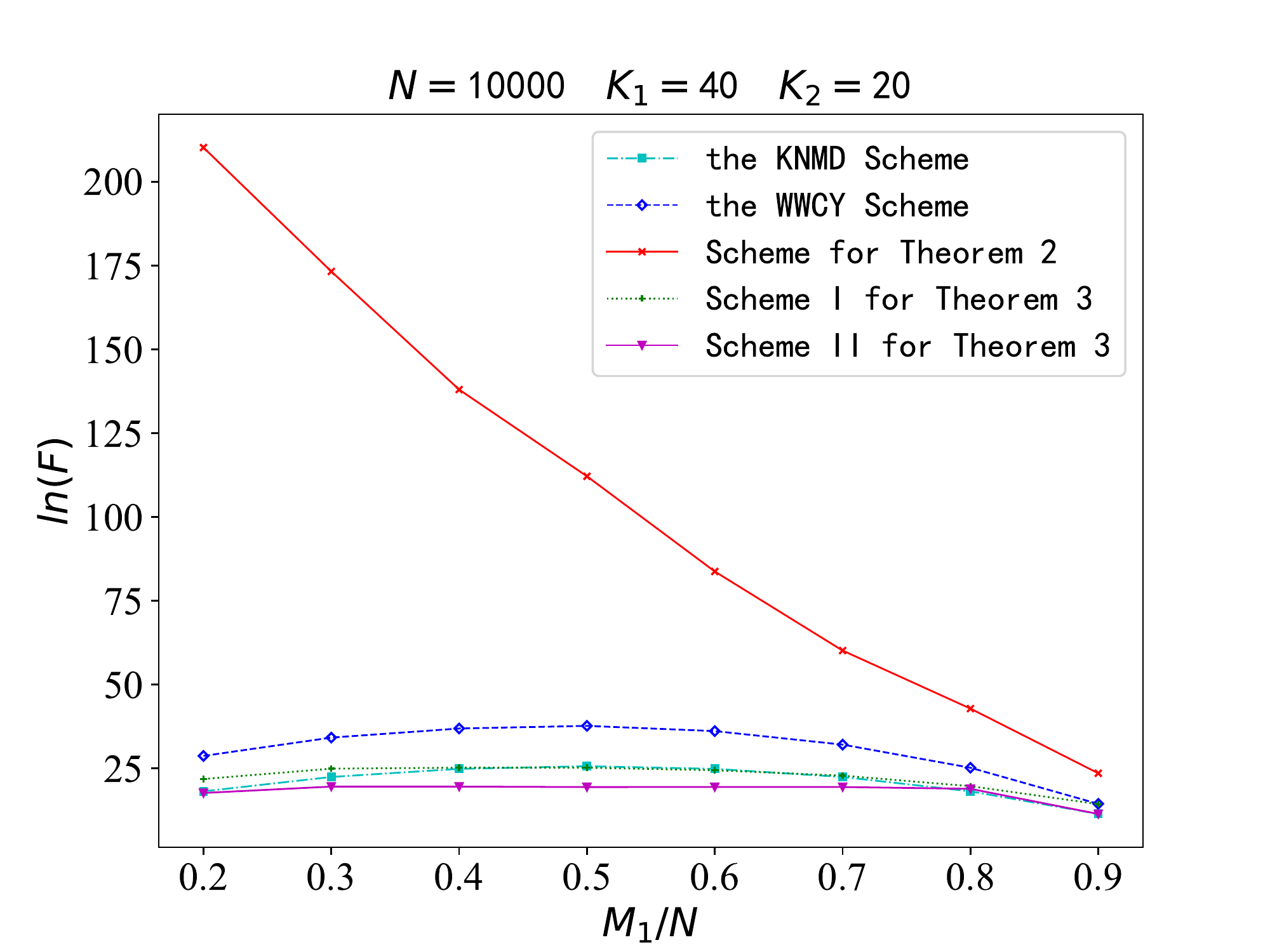}
\label{subfig3}
}
\caption{Performance comparison for a  $(K_1,K_2;M_1,M_2;N)$ caching system with $N=10000$, $K_1=40$, $K_2=20$.}
\label{per-analisis}
\end{figure}

Finally we take a numerical comparision to further show the performance of our schemes. Note that when the file number is very large, the memory ratios in \eqref{eqRatioThm2_1} and \eqref{eqRatioThm2_2} approximately equals to \eqref{eqRatioThm22 1} and \eqref{eqRatioThm22 2} respectively, which means the scheme and SP-scheme realized by the same HPDA have the same memory ratios, subpacketization and transmission loads $R_1$, $R_2$. In Fig. \ref{per-analisis}, we compare the following schemes: 1) the KNMD scheme  \cite{KNMD}; 2) the  WWCY scheme  \cite{WWCY}; 3) the Scheme for Theorem \ref{th-2};  4) Scheme I for Theorem \ref{th-3}; 5)  Scheme II for Theorem \ref{th-3}.  Note that we can also design our new hybrid schemes  like  the previous works \cite{KNMD,WWCY}, which  divide the system into two subsystems  with splitting parameters ($\alpha, \beta$), and run  the proposed schemes in the first subsystem, and the MN scheme in the second subsystem. Since the optimal $\alpha$ and $\beta$ are hard to determine due to a tradeoff between $R_1$ and $R_2$, and the second subsystem  totally ignores mirror sites'  caching abilities, we only focus on schemes working in the first subsystem, i.e., compare all schemes with  $\alpha=\beta=1$. In addition, due to the limitation on memory ratios \eqref{eqRatioThm2_1} and \eqref{eqRatioThm2_2}, it is hard to compare all schemes with general $M_1,M_2\in[0:N]$.  We thus evaluate the performance of various schemes with fixed parameters $(K_1,K_2,N)=(40,20,10000)$, and varying
parameters $(M_1,M_2)$ such that the ratios in \eqref{eqRatioThm2_1} and \eqref{eqRatioThm2_2} are satisfied.
More precisely, $M_1/N$ takes the value from $0.2$ to $0.9$ regularly with step size $0.1$, and $M_2/N$ takes the value from $0.72$ to $0.1$ (without fixed step size but on a downward trend), which satisfies memory ratios \eqref{eqRatioThm2_1} and \eqref{eqRatioThm2_2}.

Fig. \ref{subfig1} plots the rate $R_1$ versus  ${M_1}/{N}$. We can see that applying the proposed   schemes for Theorem \ref{th-2} and \ref{th-3}  can significantly reduce the transmission load $R_1$  compared to the KNMD scheme. In order to have a clear view, we draw a sketch of sub-figure in Fig. \ref{subfig1} that is without the KNMD scheme. Among all schemes, the scheme for Theorem \ref{th-2} achieves the smallest $R_1$, and the WWCY scheme achieves the second best performance. Note that the WWCY scheme is a special case of Theorem \ref{th-3} where both $\mathbf{B}$ and $\mathbf{C}$ are MN PDAs, as mentioned in Remark \ref{remark-3}. By comparing the WWCY scheme, Scheme I and Scheme II for Theorem \ref{th-3}, we can see that using different PDAs for $\mathbf{B}$ and $\mathbf{C}$ results in different transmission loads, and using MN PDA for $\mathbf{B}$ and $\mathbf{C}$ would reduce the transmission load than using other PDAs. 

Fig. \ref{subfig2} compares the rate $R_2$ versus ${M_1}/{N}$. It can be seen that the scheme for Theorem \ref{th-2} requires the largest  $R_2$. In view of  Fig. \ref{subfig1} where  the scheme for Theorem \ref{th-2} achieves the optimal $R^*_1$, we obtain that there exists a tradeoff between $R_1$ and $R_2$ as we explained in Remark \ref{remark-2}. In other words, minimizing  $R_1$ may leads to a increase on $R_2$.  Note that the WWCY scheme, the KNMD scheme and the Scheme I for Theorem \ref{th-3} achieve the same $R_2$, which is the   optimal $R^*_2$ under uncoded placement,  and the gap between the optimal rate $R^*_2$ and the rate $R_2$ of the Scheme II for Theorem \ref{th-3} is almost marginal, especially when $M_1/N$ is small.  Note that the curves in Fig.  \ref{subfig2}   show  that $R_2$ increases with   ${M_1}/{N}$. This is because $R_2$ in general decreases with ${M_2}/{N}$, while ${M_2}/{N}$ increases due to  the relation $M_1/N+M_2/N\approx t/K$ indicated by \eqref{eqRatioThm2_1} and \eqref{eqRatioThm2_2}.

Fig. \ref{subfig3} demonstrates the subpacketization of various schemes. It can be seen that the Scheme for Theorem  \ref{th-2}, which achieves the minimum $R_1$, requires the highest subpacketization, and the subpacketization decreases almost linearly with $M_1/N$. The Scheme II for Theorem \ref{th-3}, which uses the proposed PDAs in \cite{YCTC} for both $\mathbf{B}$ and $\mathbf{C}$, requires the lowest subpacketization. The WWCY scheme, which uses MN PDAs for  both $\mathbf{B}$ and $\mathbf{C}$, incurs larger subpacketization than other schemes not using MN PDAs. From the above, we can conclude that by choosing different types of PDAs to construct the HPDA, one can achieve a flexible tradeoff between  the subpacketization and transmission loads.

\section{Proof OF Theorem \ref{th-1} and Corollary \ref{coro-th-1}}
\label{sec:proof-th-1}
Recall that in Example \ref{ex-2} and \ref{ex-3}, we give two concrete examples of a scheme without security and privacy and a SP-scheme obtained from the same HPDA respectively. In this section, we first introduce the SP-scheme for a hierarchical system, and meanwhile introduce the scheme without security and privacy, i.e., the Remark \ref{remark-4}, \ref{remark-5}, \ref{remark-6}, \ref{remark-8}. Both the two schemes are based on the same HPDA. Then we prove the obtained schemes satisfy the constraint of decodability in \eqref{Decodability}, and the SP-scheme further satisfies security  in \eqref{Security I} \eqref{Security II}, and privacy in \eqref{Privacy I} \eqref{Privacy II}.

Given a $(K_1,K_2;F;Z_1,Z_2;\mathcal{S}_{\text{M}}$, $\mathcal{S}_1,\ldots,\mathcal{S}_{K_1})$ HPDA $\mathbf{A}=(\mathbf{A}^{(0)},\mathbf{A}^{(1)},\ldots,\mathbf{A}^{(K_1)})$, where $\mathbf{A}^{(0)}=(a^{(0)}_{j,k_1})_{j\in[F],k_1\in [K_1]}$ and $\mathbf{A}^{(k_1)}=(a^{(k_1)}_{j,k_2})_{j\in[F],k_2\in [K_2]}$ for each $k_1\in [K_1]$, we have an $F$-division SP-scheme for a $(K_1,K_2; M_1,M_2; N)$ hierarchical caching problem which contains the following two phases.

\subsection{Placement Phase}
\label{scheme-placement}
Divide each file into $F$ equal-size packets, i.e., $W_{n}=\{W_{n,j}\ |\ j\in[F]\}$ where $n\in [N]$. Considering the security of the transmission, i.e., the constraints \eqref{Security I} and \eqref{Security II},  the server generates $|\bigcup_{k_1=1}^{K_1}\mathcal{S}_{k_1}|=S$ security vectors from $\mathbb{F}^{B/F}_q$ independently and uniformly, denoted by $\mathcal{V}=\{\mathbf{V}_s|\ s\in\bigcup_{k_1=1}^{K_1}\mathcal{S}_{k_1}\}$. In order to protect the privacy of the transmission, i.e., the constraints \eqref{Privacy I} and \eqref{Privacy II}, for each user $\text{U}_{k_1,k_2}$ the server generates a i.i.d random privacy vector $\mathbf{p}_{k_1,k_2}=(p^{(1)}_{k_1,k_2},\ldots,p^{(N)}_{k_1,k_2})\in\mathbb{F}^{N}_q$, $k_1\in[K_1]$, $k_2\in[K_2]$.
We define the public vector set as
\begin{eqnarray}
\label{def-P}
  \mathcal{P}\triangleq\left\{\mathbf{p}_{k_1,k_2}|\ k_1\in[K_1], k_2\in[K_2]\right\},
\end{eqnarray}
where all  $\{\mathbf{p}_{k_1,k_2}\}$ are uniformly and independently generated as follows
\begin{IEEEeqnarray}{rCl}\label{pk1k2define}
\mathbf{p}_{k_1,k_2}:=({p}^{(1)}_{k_1,k_2},\ldots, {p}^{(N)}_{k_1,k_2})\sim \text{Unif}\big\{(x_1,\ldots,x_N)\in\mathbb{F}_q^N:\sum_{n\in[N]}x_n=q-1\big\}.
\end{IEEEeqnarray}
For any $\mathcal{T}_1\subseteq[K_1]$, $\mathcal{T}_2\subseteq[K_2]$, define $\mathcal{P}_{\mathcal{T}_1,\mathcal{T}_2}\triangleq\{\mathbf{p}_{k_1,k_2}| \ k_1\in\mathcal{T}_1, k_2\in\mathcal{T}_2\}$ and $\mathcal{P}_{\mathcal{T}_1,[K_2]}$ is shortened by $\mathcal{P}_{\mathcal{T}_1}$.
For any vector $\mathbf{v}=(v_1,\ldots,v_N)$ of $N$-length, we use the following notation to denote a linear combination of packets.
\begin{eqnarray}
\label{def-request-packet}
    L_{\mathbf{v},j} &:=& \sum\limits^{N}\limits_{n=1}v_n\cdot W_{n,j},\ \ \ \ \ \ \ \ \ j\in[F].
\end{eqnarray}
According to the HPDA $\mathbf{A}$, the placement strategies for the mirror sites and users are as follows.
\begin{itemize}
\item {\bf Placement for mirror sites}: Each mirror site $\text{M}_{k_1}$, $k_1\in[K_1]$ caches the packets of all files indexed by the row labels of star entries in the $k_1$-th column of $\mathbf{A}^{(0)}$. For the security of the delivery in the second layer, each mirror site $\text{M}_{k_1}$, $k_1\in[K_1]$ caches some security vectors $\mathbf{V}_s$, $s\in\mathcal{S}_{k_1}\bigcap\mathcal{S}_{\text{M}}$. Then the cache contents of mirror site $\text{M}_{k_1}$ can be written as
\begin{subequations}
\label{mirror-cache}
\begin{eqnarray}
  \mathcal{Z}_{k_1} = &\left\{W_{n,j}\ |\ a^{(0)}_{j,k_1}=*, n\in [N], j\in[F]\right\}\label{cach_mir_uncoded}\\
  &\bigcup\left\{\mathbf{V}_s\ |\ s\in\mathcal{S}_{k_1}\bigcap\mathcal{S}_{\text{M}}\right\}.\label{cach_mir_coded}
\end{eqnarray}
\end{subequations}
\item {\bf Placement for users}: Each user $\mathbf{U}_{k_1,k_2}$, $k_1\in[K_1]$, $k_2\in[K_2]$ caches the packets of all files indexed by the row labels of star entries in the $k_2$-th column of $\mathbf{A}^{(k_1)}$. In addition, each user caches some coded contents by utilizing its privacy vector and some security vectors, and the rule is that  $\mathbf{U}_{k_1,k_2}$ caches $\mathbf{V}_s+ L_{\mathbf{p}_{k_1,k_2},j}$ for each $a^{(k_1)}_{j,k_2}=s$, where $k_1\in[K_1]$, $k_2\in[K_2]$, $j\in[F]$. Then the cache contents of user $\mathbf{U}_{k_1,k_2}$ can be written as
\begin{subequations}
\label{user-cache}
\begin{eqnarray}
  \widetilde{\mathcal{Z}}_{k_1,k_2} = &\left\{W_{n,j}\ |\ a^{(k_1)}_{j,k_2}=*, n\in [N], j\in[F]\right\}\label{cach_user_uncoded}\\
  &\bigcup\left\{\mathbf{V}_s+ L_{\mathbf{p}_{k_1,k_2},j}|\ a^{(k_1)}_{j,k_2}=s,j\in[F]\right\}.\label{cach_user_coded}
\end{eqnarray}
\end{subequations}
\end{itemize}
By B$1$, B$2$ in Definition \ref{def-H-PDA} and \eqref{mirror-cache}, \eqref{user-cache}, we have $M_1=N\frac{Z_1}{F}+\frac{|\mathcal{S}_{\text{M}}\bigcap\mathcal{S}_{k_1}|}{F}$, $M_2=N\frac{Z_2}{F}+\frac{F-Z_2}{F}$, which satisfy the memory constraints $\frac{M_1}{N}=\frac{Z_1}{F}+\frac{|\mathcal{S}_{\text{M}}\bigcap\mathcal{S}_{k_1}|}{NF}$, $\frac{M_2}{N}=\frac{Z_2}{F}+\frac{F-Z_2}{NF}$.

\begin{remark}
\label{remark-4}
\eqref{cach_mir_coded} and \eqref{cach_user_coded} are coded contents only useful for protecting the security and privacy of the delivery. If the system is without the security and privacy constraints \eqref{Security I}-\eqref{Privacy II}, mirror site $\text{M}_{k_1}$ only caches
\eqref{cach_mir_uncoded} and user $\text{U}_{k_1,k_2}$ only caches \eqref{cach_user_uncoded} as follows.
\begin{eqnarray*}
  \mathcal{Z}_{k_1} = &\left\{W_{n,j}\ |\ a^{(0)}_{j,k_1}=*, n\in [N], j\in[F]\right\},\ \ \ \
  \widetilde{\mathcal{Z}}_{k_1,k_2} = &\left\{W_{n,j}\ |\ a^{(k_1)}_{j,k_2}=*, n\in [N] j\in[F]\right\}.
\end{eqnarray*}

Consequently the memory ratios are $\frac{M_1}{N}=\frac{Z_1}{F}$ and $\frac{M_2}{N}=\frac{Z_2}{F}$.\qed
\end{remark}

\subsection{Delivery Phase}
\label{scheme-delivery}
For any demand matrix $\mathbf{D}$ and all the privacy vectors $\mathbf{p}_{k_1,k_2}$, $k_1\in[K_1]$, $k_2\in[K_2]$, the server first generates $K_1K_2$ public vectors $\mathbf{q}_{k_1,k_2}$ in the following way.
\begin{eqnarray}
\label{def-Q}
  &\mathbf{q}_{k_1,k_2}=\mathbf{p}_{k_1,k_2}+\mathbf{d}_{k_1,k_2}=\left(q^{(1)}_{k_1,k_2},\ldots,q^{(N)}_{k_1,k_2}\right)\in\mathbb{F}^{N}_q, k_1\in[K_1], k_2\in[K_2].
\end{eqnarray}
By \eqref{pk1k2define} and \eqref{def-Q}, we obtain that all vectors $\{\mathbf{q}_{k_1,k_2}\}$  are  uniformly and independently distributed over the $N-1$ dimensional subspace as follows:
\begin{IEEEeqnarray}{rCl}\label{qk1k2define}
\mathbf{q}_{k_1,k_2}:=({q}^{(1)}_{k_1,k_2},\ldots, {q}^{(N)}_{k_1,k_2})\sim \text{Unif}\big\{(x_1,\ldots,x_N)\in\mathbb{F}_q^N:\sum_{n\in[N]}x_n=0\big\}.
\end{IEEEeqnarray}   We define the public vector set as
\begin{eqnarray}
\label{definition-Q}
  \mathcal{Q}\triangleq\left\{\mathbf{q}_{k_1,k_2}|\ k_1\in[K_1], k_2\in[K_2]\right\}.
\end{eqnarray}
For any $\mathcal{T}_1\subseteq[K_1]$, $\mathcal{T}_2\subseteq[K_2]$, define $\mathcal{Q}_{\mathcal{T}_1,\mathcal{T}_2}\triangleq\{\mathbf{q}_{k_1,k_2}| \ k_1\in\mathcal{T}_1, k_2\in\mathcal{T}_2\}$ and $\mathcal{Q}_{\mathcal{T}_1,[K_2]}$ is shortened by $\mathcal{Q}_{\mathcal{T}_1}$.
The server sends $\mathcal{Q}$ to all mirror sites. There are two types of message transmitted by the sever and mirror sites.
\begin{itemize}
\item \textbf{The messages sent by the server:} For each $s\in \left(\bigcup_{k_1=1}^{K_1}\mathcal{S}_{k_1}\right)\setminus\mathcal{S}_{\text{M}}$, the server transmits the following coded signal $X^{\text{server}}_s$ to all the mirror sites.
    \begin{eqnarray}
    \label{signal-server}
     X^{\text{server}}_s=\mathbf{V}_s+ \!\!\!\!\!\!\!\sum\limits_{
    a^{(k_1)}_{j,k_2}=s,j\in[F]\atop
    k_1\in[K_1],k_2\in[K_2]
    }\!\!\!\!\!\!\!L_{\mathbf{q}_{k_1,k_2},j}.
    \end{eqnarray}
    Define $\mathcal{X}^{\text{server}}=\left\{X^{\text{server}}_s|\ s\in \left(\bigcup_{k_1=1}^{K_1}\mathcal{S}_{k_1}\right)\setminus\mathcal{S}_{\text{M}}\right\}$.

\item \textbf{The messages sent by mirror site:} For each $s\in \mathcal{S}_{k_1}\setminus\mathcal{S}_{\text{M}}$, mirror site $\text{M}_{k_1}$ sends the following processed coded signal $X^{\text{mirror}}_{k_1,s}$ by canceling all the packets in $X^{\text{server}}_{s}$ which have not been requested by any user in $\mathcal{U}_{k_1}$ and have been cached by $\text{M}_{k_1}$.
    \begin{eqnarray}
    \label{signal-server-mirror}
    X^{\text{mirror}}_{k_1,s} &=& X^{\text{server}}_s-\!\!\!\!\!\!\!\sum\limits_{
    a^{(k'_1)}_{j',k'_2}=s,a^{(0)}_{j',k_1}=*\atop
    k'_2\in[K_2],j'\in[F],k'_1\in [K_1]\backslash\{k_1\}
    }\!\!\!\!\!\!\! L_{\mathbf{q}_{k'_1,k'_2},j'}\nonumber \\
    &=&\mathbf{V}_s+ \!\!\!\!\!\!\!\sum\limits_{
    a^{(k_1)}_{j,k_2}=s,j\in[F]\atop
    k_1\in[K_1],k_2\in[K_2] }\!\!\!\!\!\!\!
    L_{\mathbf{q}_{k_1,k_2},j}-\!\!\!\!\!\!\!\sum\limits_{
     a^{(k'_1)}_{j',k'_2}=s,a^{(0)}_{j',k_1}=*\atop
     k'_2\in[K_2], j'\in[F],k'_1\in [K_1]\backslash\{k_1\}
    }\!\!\!\!\!\!\! L_{\mathbf{q}_{k'_1,k'_2},j'}.
    \end{eqnarray}
    For each $s'\in \mathcal{S}_{k_1}\bigcap\mathcal{S}_{\text{M}}$, mirror site $\text{M}_{k_1}$ directly generates the following coded signals $X^{\text{mirror}}_{k_1,s'}$ similar to the delivery strategy realized by PDA $\mathbf{A}^{(k_1)}$.
    \begin{eqnarray}
    \label{signal-mirror}
    X^{\text{mirror}}_{k_1,s'} =\mathbf{V}_{s'}+ \!\!\!\!\!\!\!\sum_{a^{(k_1)}_{j,k_2}=s',j\in[F],k_2\in[K_2]}\!\!\!\!\!\!\!L_{\mathbf{q}_{k_1,k_2},j}.
    \end{eqnarray}
    Define $\mathcal{X}^{\text{mirror}}_{k_1} \triangleq\{X^{\text{mirror}}_{k_1,s}\ |\ s\in\mathcal{S}_{k_1}\setminus\mathcal{S}_{\text{M}}\}\bigcup
    \{X^{\text{mirror}}_{k_1,s'}\ |\  s'\in\mathcal{S}_{k_1}\bigcap\mathcal{S}_{\text{M}}\}$, $\mathfrak{X}^{\text{mirror}}_{\mathcal{T}_1}\triangleq\{\mathcal{X}^{\text{mirror}}_{k_1}|\ k_1\in\mathcal{T}_1\}$.
\end{itemize}

\begin{remark}
\label{remark-5}
If the system is without the security and privacy constraints \eqref{Security I}-\eqref{Privacy II}, then we delete the security vectors $\mathbf{V}_s$ where $s\in\bigcup^{K_1}_{k_1=1}\mathcal{S}_{k_1}$ and the public vectors $\mathbf{q}_{k_1,k_2}$ where $k_1\in[K_1]$, $k_2\in[K_2]$, and \eqref{signal-server}-\eqref{signal-mirror} can be written as follows respectively.
    \begin{eqnarray}
    X^{\text{server}}_s&=&\!\!\!\!\!\!\!\sum_{
    a^{(k_1)}_{j,k_2}=s,j\in[F]\atop
    k_1\in[K_1],k_2\in[K_2] }\!\!\!\!\!\!\!L_{\mathbf{q}_{k_1,k_2},j},\ s\in \left(\bigcup_{k_1=1}^{K_1}\mathcal{S}_{k_1}\right)\setminus\mathcal{S}_{\text{M}}\label{signal-server-ori}\\[0.2cm]
    X^{\text{mirror}}_{k_1,s} &=& X^{\text{server}}_s-\!\!\!\!\!\!\!\sum\limits_{
     a^{(k'_1)}_{j',k'_2}=s,a^{(0)}_{j',k_1}=*\atop
     k'_2\in[K_2], j'\in[F],k'_1\in [K_1]\backslash\{k_1\}
    }\!\!\!\!\!\!\!L_{\mathbf{d}_{k'_1,k'_2},j'}\nonumber\\
    &=&\!\!\!\!\!\!\!\sum_{
    a^{(k_1)}_{j,k_2}=s,j\in[F]\atop
    k_1\in[K_1],k_2\in[K_2]}\!\!\!\!\!\!\!L_{\mathbf{d}_{k_1,k_2},j}-\!\!\!\!\!\!\!\sum\limits_{
     a^{(k'_1)}_{j',k'_2}=s,a^{(0)}_{j',k_1}=*\atop
     k'_2\in[K_2], j'\in[F],k'_1\in [K_1]\backslash\{k_1\}
    }\!\!\!\!\!\!\!L_{\mathbf{d}_{k'_1,k'_2},j'},\ s\in \mathcal{S}_{k_1}\setminus\mathcal{S}_{\text{M}}\label{signal-server-mirror-ori}\\[0.2cm]
    X^{\text{mirror}}_{k_1,s'}&=&\!\!\!\!\!\!\!\sum_{a^{(k_1)}_{j,k_2}=s',j\in[F],k_2\in[K_2]}\!\!\!\!\!\!\!L_{\mathbf{d}_{k_1,k_2},j},\ s'\in \mathcal{S}_{k_1}\bigcap\mathcal{S}_{\text{M}}.\label{signal-mirror-ori}
\end{eqnarray}
\end{remark}

\subsection{Decodability}
\label{scheme-decodability}
Recall that $a^{(k_1)}_{j,k_2}=*$ represents user $\text{U}_{k_1,k_2}$ has cached all the files packets indexed by $j$ for any given integers $j\in[F]$, $k_1\in[K_1]$ and $k_2\in[K_2]$. From \eqref{def-request-packet}, user  $\text{U}_{k_1,k_2}$ can also obtain the coded packet $L_{\mathbf{d}_{k_1,k_2},j}$. So it is sufficient to show that each user $\text{U}_{k_1,k_2}$ can obtain all the coded packets  $L_{\mathbf{d}_{k_1,k_2},j}$ where $a^{(k_1)}_{j,k_2}\neq *$, $j\in[F]$, $k_1\in[K_1]$ and $k_2\in[K_2]$ by its received coded packets $\mathcal{X}^{\text{mirror}}_{k_1}$, and its cached packets.

First, when $a^{(k_1)}_{j,k_2}=s\in \mathcal{S}_{k_1}\setminus\mathcal{S}_{\text{M}}$, we will show that user $\text{U}_{k_1,k_2}$ can decode its desired linear combination packets $L_{\mathbf{d}_{k_1,k_2},j}$ from $X^{\text{mirror}}_{k_1,s}$ according to \eqref{signal-server-mirror} as follows.
\begin{subequations}
\label{sig-mirror1}
\begin{IEEEeqnarray}{rCl}
  X^{\text{mirror}}_{k_1,s}&=&\mathbf{V}_s+ \!\!\!\!\!\!\!\sum\limits_{
    a^{(k'_1)}_{j',k'_2}=s,j'\in[F]\atop
    k'_1\in[K_1],k'_2\in[K_2] }\!\!\!\!\!\!\!
    L_{\mathbf{q}_{k'_1,k'_2},j'}-\!\!\!\!\!\!\!\sum\limits_{
     a^{(k''_1)}_{j'',k''_2}=s,a^{(0)}_{j'',k_1}=*\atop
     k''_2\in[K_2], j''\in[F],k''_1\in [K_1]\backslash\{k_1\}
    }\!\!\!\!\!\!\! L_{\mathbf{q}_{k''_1,k''_2},j''}\nonumber\\
  &\stackrel{(a)}=&\mathbf{V}_{s}+L_{\mathbf{q}_{k_1,k_2},j}+ \!\!\!\!\!\!\!\sum_{
     a^{(k'_1)}_{j',k'_2}=s,j'\in[F]\atop
     (k'_1,k'_2)\in\left([K_1]\times[K_2]\right)\setminus\{(k_1,k_2)\}
    }\!\!\!\!\!\!\!L_{\mathbf{q}_{k'_1,k'_2},j'}-\!\!\!\!\!\!\!\sum\limits_{
     a^{(k''_1)}_{j'',k''_2}=s,a^{(0)}_{j'',k_1}=*\atop
     k''_2\in[K_2], j''\in[F],k''_1\in [K_1]\backslash\{k_1\}
    }\!\!\!\!\!\!\!L_{\mathbf{q}_{k''_1,k''_2},j''}\label{sig-mirror1-1}\\[0.2cm]
  &\stackrel{(b)}=&\mathbf{V}_{s}\!+\!L_{\mathbf{d}_{k_1,k_2},j}+ L_{\mathbf{p}_{k_1,k_2},j}\!+\! \!\!\!\!\!\!\!\sum_{
     a^{(k'_1)}_{j',k'_2}=s,j'\in[F]\atop
     (k'_1,k'_2)\in\left([K_1]\times[K_2]\right)\setminus\{(k_1,k_2)\}
    }\!\!\!\!\!\!\!L_{\mathbf{q}_{k'_1,k'_2},j'}-\!\!\!\!\!\!\!\!\!\!\!\!\!\sum\limits_{
     a^{(k''_1)}_{j'',k''_2}=s,a^{(0)}_{j'',k_1}=*\atop
     k''_2\in[K_2], j''\in[F],k''_1\in [K_1]\backslash\{k_1\}
    }\!\!\!\!\!\!\!\!\!\!\!\!\!L_{\mathbf{q}_{k''_1,k''_2},j''}\label{sig-mirror1-2}\\[0.2cm]
  &=&\mathbf{V}_{s}+L_{\mathbf{d}_{k_1,k_2},j}+ L_{\mathbf{p}_{k_1,k_2},j}+\!\!\!\!\!\!\!\sum\limits_{
     a^{(k'_1)}_{j',k'_2}=s,a^{(k_1)}_{j',k_2}=*,j'\in[F]\atop
    (k'_1,k'_2)\in([K_1]\times[K_2])\setminus\{(k_1,k_2)\}
    }\!\!\!\!\!\!\!L_{\mathbf{q}_{k'_1,k'_2},j'}\label{sig-mirror1-3}\\[0.2cm]
  &=&\underbrace{L_{\mathbf{d}_{k_1,k_2},j}}_{\text{the desired contents by $\text{U}_{k_1,k_2}$ }}\!+\!\underbrace{(\mathbf{V}_{s}\!+\!L_{\mathbf{p}_{k_1,k_2},j})}_{\text{the coded contents cached by $\text{U}_{k_1,k_2}$}}+\underbrace{\!\!\!\!\!\!\!\!\!\!\sum\limits_{
     a^{(k'_1)}_{j',k'_2}=s,a^{(k_1)}_{j',k_2}=*,j'\in[F]\atop
    (k'_1,k'_2)\in([K_1]\times[K_2])\setminus\{(k_1,k_2)\}
    }\!\!\!\!\!\!\!\!\!\!L_{\mathbf{q}_{k'_1,k'_2},j'}}_{\text{file packets cached by $\text{U}_{k_1,k_2}$}}.\label{sig-mirror1-4}
\end{IEEEeqnarray}
\end{subequations}
Here \eqref{sig-mirror1-1} can be directly obtained by the condition  B2 of HPDA, i.e., there is only one $s$ in the $k_2$-th column of $\mathbf{A}^{(k_1)}$. Recall that $L_{\mathbf{q}_{k_1,k_2},j}=L_{\mathbf{d}_{k_1,k_2},j}+L_{\mathbf{p}_{k_1,k_2},j}$, where $k_1\in[K_1]$, $k_2\in[K_2]$, $j\in[F]$ by \eqref{def-request-packet} and \eqref{def-Q}. Then \eqref{sig-mirror1-2} is derived. By the condition B$4$ of Definition \ref{def-H-PDA}, if $a^{(k_1)}_{j,k_2}=a^{(k'_1)}_{j',k'_2}=s$, where $(k'_1,k'_2)\in\left([K_1]\times[K_2]\right)\setminus\{(k_1,k_2)\}$, we have two situations. The first is $a^{(k_1)}_{j',k_2}=*$, and the second is  $a^{(k_1)}_{j',k_2}\neq*$ and $a^{(0)}_{j',k_1}=*$ which means that $\text{U}_{k_1,k_2}$ and $\text{M}_{k_1}$ have together cached all the packets indicated by $s$ except for the required packets. So after the the subtracting of the packets which are not desired by $\text{U}_{k_1,k_2}$ and cached by $\text{M}_{k_1}$, i.e., deleting the last term from the forth term in \eqref{sig-mirror1-2}, the remained packets are all cached by $\text{U}_{k_1,k_2}$, i.e., the last term of \eqref{sig-mirror1-3}. Finally, since each user knows the public vectors, with the help of $\widetilde{Z}_{k_1,k_2}$, $\text{U}_{k_1,k_2}$ can retrieve its desired linear combination packet $L_{\mathbf{d}_{k_1,k_2},j}$ from \eqref{sig-mirror1-4}.

Second when $a^{(k_1)}_{j,k_2}=s'\in\mathcal{S}_{k_1}\bigcap\mathcal{S}_{\text{M}}$, we will show that user $\text{U}_{k_1,k_2}$ can decode its desired linear combination packets $L_{\mathbf{d}_{k_1,k_2},j}$ from $X^{\text{mirror}}_{k_1,s'}$ according to \eqref{signal-mirror} as follows.
\begin{subequations}
\label{sig-mirror2}
\begin{IEEEeqnarray}{rCl}
  X^{\text{mirror}}_{k_1,s'}&=&\mathbf{V}_{s'}+ \!\!\!\!\!\!\!\sum_{a^{(k_1)}_{j',k'_2}=s',j'\in[F],k'_2\in[K_2]}\!\!\!\!\!\!\!L_{\mathbf{q}_{k_1,k'_2},j'}\nonumber\\
  &\stackrel{(a)}=&\mathbf{V}_{s'}+L_{\mathbf{q}_{k_1,k_2},j}+ \!\!\!\!\!\!\!\sum_{a^{(k_1)}_{j',k'_2}=s',a^{(k_1)}_{j',k_2}=*\atop j'\in[F],k'_2\in[K_2]\setminus\{k_2\}}\!\!\!\!\!\!\!L_{\mathbf{q}_{k_1,k'_2},j'}\label{sig-mirror2-1}\\[0.2cm]
  &=&\mathbf{V}_{s'}+L_{\mathbf{d}_{k_1,k_2},j}+ L_{\mathbf{p}_{k_1,k_2},j}+\!\!\!\!\!\!\!\sum_{a^{(k_1)}_{j',k'_2}=s',a^{(k_1)}_{j',k_2}=*\atop j'\in[F],k'_2\in[K_2]\setminus\{k_2\}}\!\!\!\!\!\!\!L_{\mathbf{q}_{k_1,k'_2},j'}\nonumber\\[0.2cm]
  &=&\underbrace{L_{\mathbf{d}_{k_1,k_2},j}}_{\text{the desired contents}\atop\text{by $\text{U}_{k_1,k_2}$ }}+\underbrace{(\mathbf{V}_{s'}+L_{\mathbf{p}_{k_1,k_2},j})}_{\text{the coded contents}\atop\text{cached by $\text{U}_{k_1,k_2}$ }}+\underbrace{\!\!\!\!\!\!\!\sum_{a^{(k_1)}_{j',k'_2}=s',a^{(k_1)}_{j',k_2}=*\atop j'\in[F],k'_2\in[K_2]\setminus\{k_2\}}\!\!\!\!\!\!\!L_{\mathbf{q}_{k_1,k'_2},j'}}_{\text{file packets cached by $\text{U}_{k_1,k_2}$}}.\label{sig-mirror2-2}
\end{IEEEeqnarray}
\end{subequations}
Similarly \eqref{sig-mirror2-1} can be directly obtained by the condition B2 of HPDA. In addition,
the generation of signal $X^{\text{mirror}}_{k_1,s'}$, $s'\in\mathcal{S}_{k_1}\bigcap\mathcal{S}_{\text{M}}$ is similar to the delivery strategy realized by PDA $\mathbf{A}_{(k_1)}$. According to the condition C$3$ of definition \ref{def-PDA}, if $a^{(k_1)}_{j,k_2}=a^{(k_1)}_{j',k'_2}=s'$, where $k'_2\in[K_2]\setminus\{k_2\}$, $j'\in[F]$ we have $a^{(k_1)}_{j',k_2}=*$, which means that $\text{U}_{k_1,k_2}$ has cached all the packets indexed by $s'$ except for the desired packets. Finally, since each user knows the public vectors, with the help of $\widetilde{Z}_{k_1,k_2}$, $\text{U}_{k_1,k_2}$ can retrieve its desired linear combination packet $L_{\mathbf{d}_{k_1,k_2},j}$ from \eqref{sig-mirror2-2}.

\begin{remark}
\label{remark-6}
When we delete the security vectors and privacy vectors in \eqref{sig-mirror1-4} and \eqref{sig-mirror2-2}, i.e., the coded contents, it is easy to check that each user can also decode its required linear combination packet. That is the scheme without security and privacy constraints \eqref{Security I}-\eqref{Privacy II}.
\end{remark}

\subsection{Security}
We prove the security that the transmitted signals at the server and mirror sites can not be used to decode any content of the library and users' request, i.e.,
$I\left(\mathcal{D},\mathcal{W}; X^{\text{server}}\right)=0$ and $I\left(\mathcal{D},\mathcal{W}; X^{\text{mirror}}_{1},\ldots,X^{\text{mirror}}_{K_1}\right)=0.$
The proof follows similar to that in \cite{YDT}. we have
\begin{IEEEeqnarray*}{rCl}
I\left(\mathcal{D},\mathcal{W}; X^{\text{server}}\right)&\stackrel{(a)}=&I\left(\mathcal{D},\mathcal{W};\mathcal{Q},\mathcal{X}^\text{server}\right)\nonumber\\
&=& ~ I\left(\mathcal{D},\mathcal{W};\mathcal{Q}\right)+I\left(\mathcal{D},\mathcal{W};\mathcal{X}^\text{server}|\mathcal{Q}\right)\nonumber\\
&\stackrel{(b)}=& ~ 0 +I\left(\mathcal{D},\mathcal{W};\mathcal{X}^\text{server}|\mathcal{Q}\right)\\
&\stackrel{(c)}=& ~ 0,
\end{IEEEeqnarray*}
 where (a) follows from \eqref{signal-server}, (b) holds because $\mathcal{Q}$ is independent of $(\mathcal{D},\mathcal{W})$ as $\mathbf{p}_{k_1,k_2}$ where $k_1\in[K_1]$, $k_2\in[K_2]$  are independently and uniformly distributed over $\mathbb{F}_{q}^{N}$, and (c) holds because signals in $\mathcal{X}^\text{server}$ are independent of $(\mathcal{D},\mathcal{W},\mathcal{Q})$ as signals in $\mathcal{X}^\text{server}$ are independently and uniformly distributed over $\mathbb{F}_{q}^{B/F}$.  Moreover,  we have
 \begin{IEEEeqnarray*}{rCl}
I\left(\mathcal{D},\mathcal{W};X^{\text{mirror}}_{1},\ldots,X^{\text{mirror}}_{K_1}\right)
&\!\stackrel{(a)}=\!&I\left(\mathcal{D},\mathcal{W};\mathcal{Q},\{\mathcal{X}^{\text{mirror}}_{k_1}\}_{k_1\in[K_1]}\right)\nonumber\\
&=& ~  0+I\left(\mathcal{D},\mathcal{W};\{\mathcal{X}^{\text{mirror}}_{k_1}\}_{k_1\in[K_1]}|\mathcal{Q}\right)\\
&\stackrel{(b)}=& 0,
\end{IEEEeqnarray*}
  where (a) follows from \eqref{signal-server-mirror} and \eqref{signal-mirror} and (b)  holds because $\{\mathcal{X}^{\text{mirror}}_{k_1}\}_{k_1\in[K_1]}$ are independent of $(\mathcal{D},\mathcal{W},\mathcal{Q})$ as $\mathcal{X}^\text{server}$ are independently and uniformly distributed over $\mathbb{F}_{q}^{B/F}$.

\label{scheme-security}

\subsection{Privacy}
\label{scheme-privacy}
We first prove the privacy that any subset  of users' requests   can not be decoded by the remaining subset of users, even if they can collude with each other, i.e., the constraint in \eqref{Privacy II}.
\begin{IEEEeqnarray}{rCl}\label{Privacy1-1}
&&I\left(\mathcal{D}_{[K_1]\setminus\mathcal{T}_1,[K_2]\setminus\mathcal{T}_2};\mathcal{W},\{X^{\text{mirror}}_{k_1}\}_{k_1\in\mathcal{T}_1},\widetilde{\mathcal{Z}}_{\mathcal{T}_1,\mathcal{T}_2},\mathcal{D}_{\mathcal{T}_1,\mathcal{T}_2}\right)\nonumber\\
\stackrel{(a)}= ~ && I\left(\mathcal{D}_{[K_1]\setminus\mathcal{T}_1,[K_2]\setminus\mathcal{T}_2};  \mathcal{W},\mathfrak{X}^{\text{mirror}}_{\mathcal{T}_1},\widetilde{\mathcal{Z}}_{\mathcal{T}_1,\mathcal{T}_2},\mathcal{D}_{\mathcal{T}_1,\mathcal{T}_2}\right) \nonumber\\
\leq ~ && I\left(\mathcal{D}_{[K_1]\setminus\mathcal{T}_1,[K_2]\setminus\mathcal{T}_2};\mathcal{Q}, \mathcal{V}, \mathcal{W}, \mathfrak{X}^{\text{mirror}}_{\mathcal{T}_1}, \widetilde{\mathcal{Z}}_{\mathcal{T}_1,\mathcal{T}_2},\mathcal{D}_{\mathcal{T}_1,\mathcal{T}_2}\right)  \nonumber\\
\stackrel{(b)}= ~ && I\left(\mathcal{D}_{[K_1]\setminus\mathcal{T}_1,[K_2]\setminus\mathcal{T}_2};\mathcal{Q}, \mathcal{V},\mathcal{W},\widetilde{\mathcal{Z}}_{\mathcal{T}_1,\mathcal{T}_2},\mathcal{D}_{\mathcal{T}_1,\mathcal{T}_2}\right)\nonumber\\ \leq ~ &&  I\left(\mathcal{D}_{[K_1]\setminus\mathcal{T}_1,[K_2]\setminus\mathcal{T}_2};\mathcal{Q},\mathcal{V}, \mathcal{P}_{\mathcal{T}_1, \mathcal{T}_2},\mathcal{W},\widetilde{\mathcal{Z}}_{\mathcal{T}_1,\mathcal{T}_2},\mathcal{D}_{\mathcal{T}_1,\mathcal{T}_2}\right) \nonumber\\
\stackrel{(c)}= ~ &&  I\left(\mathcal{D}_{[K_1]\setminus\mathcal{T}_1,[K_2]\setminus\mathcal{T}_2};\mathcal{Q}_{[K_1]\setminus\mathcal{T}_1,[K_2]\setminus\mathcal{T}_2}, \mathcal{P}_{\mathcal{T}_1, \mathcal{T}_2},\mathcal{V},\mathcal{W},\widetilde{\mathcal{Z}}_{\mathcal{T}_1,\mathcal{T}_2},\mathcal{D}_{\mathcal{T}_1,\mathcal{T}_2}\right) \nonumber\\
\stackrel{(d)}= ~ &&  I\left(\mathcal{D}_{[K_1]\setminus\mathcal{T}_1,[K_2]\setminus\mathcal{T}_2};\mathcal{Q}_{[K_1]\setminus\mathcal{T}_1,[K_2]\setminus\mathcal{T}_2}, \mathcal{P}_{\mathcal{T}_1, \mathcal{T}_2},\mathcal{V},\mathcal{W},\mathcal{D}_{\mathcal{T}_1,\mathcal{T}_2}\right) \nonumber\\
= ~ && I\left(\mathcal{D}_{[K_1]\setminus\mathcal{T}_1,[K_2]\setminus\mathcal{T}_2};\mathcal{D}_{\mathcal{T}_1,\mathcal{T}_2},\mathcal{P}_{\mathcal{T}_1, \mathcal{T}_2},\mathcal{W},\mathcal{V}\right)+  I\left(\mathcal{D}_{[K_1]\setminus\mathcal{T}_1,[K_2]\setminus\mathcal{T}_2};\mathcal{Q}_{[K_1]\setminus\mathcal{T}_1,[K_2]\setminus\mathcal{T}_2}| \mathcal{P}_{\mathcal{T}_1, \mathcal{T}_2},\mathcal{W},\mathcal{D}_{\mathcal{T}_1,\mathcal{T}_2},\mathcal{V}\right) \nonumber\\
= ~ && 0+ H\left(\mathcal{Q}_{[K_1]\setminus\mathcal{T}_1,[K_2]\setminus\mathcal{T}_2}| \mathcal{P}_{\mathcal{T}_1, \mathcal{T}_2},\mathcal{W},\mathcal{D}_{\mathcal{T}_1,\mathcal{T}_2},\mathcal{V}\right)-H(\mathcal{Q}_{[K_1]\setminus\mathcal{T}_1,[K_2]\setminus\mathcal{T}_2}| \mathcal{P}_{\mathcal{T}_1, \mathcal{T}_2},\mathcal{W},\mathcal{D}_{\mathcal{T}_1,\mathcal{T}_2}, \mathcal{D},\mathcal{V})\nonumber\\
 \stackrel{(e)}= ~ && H\left(\mathcal{Q}_{[K_1]\setminus\mathcal{T}_1,[K_2]\setminus\mathcal{T}_2}\right)-H(\mathcal{P}_{[K_1]\setminus\mathcal{T}_1,[K_2]\setminus\mathcal{T}_2})\nonumber\\
 \stackrel{(f)}= ~ &&0,
\end{IEEEeqnarray}
where (a) holds because  \eqref{signal-server-mirror} and \eqref{signal-mirror}, (b) holds because $\mathfrak{X}^{\text{mirror}}_{\mathcal{T}_1}$  is determined by  $(\mathcal{Q},\mathcal{V},\mathcal{W})$ according to \eqref{signal-server-mirror} and \eqref{signal-mirror}, (c) is by \eqref{definition-Q}, (d) follows from \eqref{user-cache}, (e) is by the definition of $\mathbf{q}_{k_1,k_2}$ in \eqref{def-Q} and because  $\mathcal{Q}_{[K_1]\setminus\mathcal{T}_1,[K_2]\setminus\mathcal{T}_2}$ and $\mathcal{P}_{[K_1]\setminus\mathcal{T}_1,[K_2]\setminus\mathcal{T}_2}$ are independent on $(\mathcal{P}_{\mathcal{T}_1, \mathcal{T}_2},\mathcal{W},\mathcal{D}_{\mathcal{T}_1,\mathcal{T}_2},\mathcal{V})$ and $(\mathcal{P}_{\mathcal{T}_1, \mathcal{T}_2},\mathcal{W},\mathcal{D}_{\mathcal{T}_1,\mathcal{T}_2}, \mathcal{D},\mathcal{V})$, respectively, and (f) holds by \eqref{pk1k2define} and \eqref{qk1k2define}.

Besides, our scheme also guarantees the privacy that any subset  of requests $\mathcal{D}_{[K_1]\backslash\mathcal{T}_1}$  can not be known by the remaining   mirrors, even if they can collude with each other, i.e., the privacy constraint \eqref{Privacy I}.
 Following the similar proof in \eqref{Privacy1-1}, we obtain
\begin{IEEEeqnarray}{rCl}
&&I\left(\mathcal{D}_{[K_1]\backslash\mathcal{T}_1};\mathcal{W},X^{\text{server}},\mathcal{Z}_{\mathcal{T}_1},\mathcal{D}_{\mathcal{T}_1}\right)\\
&=&I\left(\mathcal{D}_{[K_1]\backslash\mathcal{T}_1};  \mathcal{W},\mathcal{V}, \mathcal{X}^{\text{server}},\mathcal{Z}_{\mathcal{T}_1},\mathcal{D}_{\mathcal{T}_1}\right) \nonumber\\
&\leq&I\left(\mathcal{D}_{[K_1]\backslash\mathcal{T}_1};\mathcal{Q}, \mathcal{W},\mathcal{V}, \{X^{\text{server}}_{s}\}_{ s\in\mathcal{S}_{k_1}},\mathcal{Z}_{\mathcal{T}_1},\mathcal{D}_{\mathcal{T}_1}\right) \nonumber\\
&=&I\left(\mathcal{D}_{[K_1]\backslash\mathcal{T}_1};\mathcal{Q}, \mathcal{V},\mathcal{W},\mathcal{Z}_{\mathcal{T}_1},\mathcal{D}_{\mathcal{T}_1}\right)  \nonumber\\
&\leq&I\left(\mathcal{D}_{[K_1]\backslash\mathcal{T}_1};\mathcal{Q},\mathcal{V},\mathcal{P}_{\mathcal{T}_1},\mathcal{W},\mathcal{Z}_{\mathcal{T}_1},\mathcal{D}_{\mathcal{T}_1}\right) \nonumber\\
&=&I\left(\mathcal{D}_{[K_1]\backslash\mathcal{T}_1};\mathcal{Q}_{[K_1]\backslash\mathcal{T}_1},\mathcal{P}_{\mathcal{T}_1},\mathcal{W},\mathcal{V},\mathcal{Z}_{\mathcal{T}_1},\mathcal{D}_{\mathcal{T}_1}\right) \nonumber\\
&=&I\left(\mathcal{D}_{[K_1]\backslash\mathcal{T}_1};\mathcal{D}_{\mathcal{T}_1},\mathcal{P}_{\mathcal{T}_1},\mathcal{W},\mathcal{V}\right)+  I\left(\mathcal{D}_{[K_1]\backslash\mathcal{T}_1};\mathcal{Q}_{[K_1]\backslash\mathcal{T}_1}| \mathcal{P}_{\mathcal{T}_1},\mathcal{W},\mathcal{D}_{\mathcal{T}_1},\mathcal{V}\right) \nonumber\\
&=&0+ H\left(\mathcal{Q}_{[K_1]\backslash\mathcal{T}_1}| \mathcal{P}_{\mathcal{T}_1},\mathcal{W},\mathcal{D}_{\mathcal{T}_1},\mathcal{V}\right)-H(\mathcal{Q}_{[K_1]\backslash\mathcal{T}_1}| \mathcal{P}_{\mathcal{T}_1},\mathcal{W},\mathcal{D}_{\mathcal{T}_1},\mathcal{V})\nonumber\\
&=&H\left(\mathcal{Q}_{[K_1]\backslash\mathcal{T}_1}\right)-H(\mathcal{P}_{[K_1]\backslash\mathcal{T}_1}) = ~ 0.
\end{IEEEeqnarray}

\begin{remark}
\label{remark-7}
For the desired packets that have been cached by mirror site, they are sent directly from the mirror site, which results in the mirror site knowing the users' demand. We can also protect the privacy of $\mathcal{U}_{k_1}$ against $\text{M}_{k_1}$. That is, let server send \eqref{signal-mirror} and each mirror site only forward the signals. Obviously this causes a larger transmission load $R_1$.
\end{remark}

\subsection{Performance}
\label{scheme-performance}
From the Placement phase, each file is partitioned into $F$ packets with packet size $B/F$. According to the delivery phase, the number of coded packets generated by the server is $|\bigcup_{k_1=1}^{K_1}\mathcal{S}_{k_1}\setminus\mathcal{S}_{\text{M}}|$, and the number of coded packets transmitted by the mirror site $\text{M}_{k_1}$ is $|\mathcal{S}_{k_1}|$. Then we have
\begin{eqnarray}
  R_1 &=& \frac{|\bigcup_{k_1=1}^{K_1}\mathcal{S}_{k_1}|-|\mathcal{S}_{\text{M}}|}{F}, \label{eq-R_1}\\
  R_2 &=& \max_{k_1\in[K_1]}\left\{\  \frac{\mid\mathcal{S}_{k_1}\mid}{F}\ \right\}.\label{eq-R_2}
\end{eqnarray}

\begin{remark}
\label{remark-8}
    The number of coded packets in the transmitting signals \eqref{signal-server-ori}, \eqref{signal-server-mirror-ori} and \eqref{signal-mirror-ori} are the same as \eqref{signal-server}, \eqref{signal-server-mirror} and \eqref{signal-mirror} respectively, which indicates that the scheme without security and privacy also achieves the same transmission loads $R_1$ in \eqref{eq-R_1} and $R_2$ in \eqref{eq-R_2}.
\end{remark}

\section{Proof of Theorem \ref{th-2}}
\label{sec:proof-th-2}
 In this section, we first describe how to construct an HPDA by the grouping method based on a MN PDA, and then we prove the obtained HPDA leads to a hierarchical coded caching scheme achieving the performance in Theorem \ref{th-2}.
\subsection{The Construction of Grouping Method}
For the sake of convenience, we use a set $\mathcal{T}\in {[K_1K_2]\choose t}$ to represent the row index of the MN PDA defined in Construction \ref{con-MN} and the constructed HPDA. Given a $(K,F,Z,S)=$ $(K_1K_2,{K_1K_2\choose t},{K_1K_2-1\choose t-1},{K_1K_2\choose t+1})$ MN PDA $\mathbf{B}=(b_{\mathcal{T},k})_{\mathcal{T}\in {[K_1K_2]\choose t},k\in[K_1K_2]}$, where $t\in[K_2:K_1K_2]$, we show how to construct a $(K_1,K_2$; $F={K_1K_2\choose t}$; $Z_1={K_1K_2-K_2\choose t-K_2}$, $Z_2={K_1K_2-1\choose t-1}-{K_1K_2-K_2\choose t-K_2}$; $\mathcal{S}_{\text{M}},\mathcal{S}_1$, $\ldots$, $\mathcal{S}_{K_1})$ HPDA  $\mathbf{A}=\left(\mathbf{A}^{(0)}\right.$, $\mathbf{A}^{(1)}$, $\ldots$, $\left.\mathbf{A}^{(K_1)}\right)$, where
\begin{IEEEeqnarray*}{rCl}
&& \mathbf{A}^{(0)}=(a^{(0)}_{\mathcal{T},k_1})_{\mathcal{T}\in {[K_1K_2]\choose t},k_1\in [K_1]},~a^{(0)}_{\mathcal{T},k_1}\in\{*,null\},\\
&& \mathbf{A}^{(k_1)}=(a^{(k_1)}_{\mathcal{T},k_2})_{\mathcal{T}\in {[K_1K_2]\choose t},k_2\in[(k_1-1)K_2+1:k_1K_2]}, ~a^{(k_1)}_{\mathcal{T},k_2}\in\{*\}\cup \mathcal{S}_{k_1},\ k_1\in [K_1].
\end{IEEEeqnarray*}
The integer sets $\mathcal{S}_{\text{M}}$, $\mathcal{S}_{k_1}$ are listed in \eqref{eq-alphabet-m} and \eqref{eq-S_k_1} respectively. The constructions of $\mathbf{A}^{(0)}$ and  $(\mathbf{A}^{(1)}, \ldots, \mathbf{A}^{(K_1)} )$ are described  as the following three steps.
\begin{itemize}
\item{\bf Step 1.} Construction of $\mathbf{A}^{(0)}$. We first divide the MN PDA $\mathbf{B}$ into equal $K_1$ parts by columns, i.e., $\mathbf{B}=\left(\mathbf{B}^{(1)}\right.$,$\ldots$, $\left.\mathbf{B}^{(K_1)}\right)$, and obviously for each $k_1\in[K_1]$, $\mathbf{B}^{(k_1)}$ is a $F\times K_2$ PDA, where $\mathbf{B}^{(k_1)}\!=\!(b^{(k_1)}_{\mathcal{T},k_2})_{\mathcal{T}\in {[K_1K_2]\choose t},k_2\in[(k_1-1)K_2+1:k_1K_2]}$.
 We construct the $F\times K_1$ array $\mathbf{A}^{(0)}$ by the following rule:
\begin{eqnarray}
\label{eq-mirror-cache1}
a^{(0)}_{\mathcal{T},k_1}=\left\{\begin{array}{cl}
                  *, & ~\text{if~} b^{(k_1)}_{\mathcal{T},k_2}=*, \begin{array}{c} \forall k_2\in[(k_1-1)K_2+1:k_1K_2]
                                  \end{array}
                                  \\
                  \text{null}, &  \text{otherwise}.
                \end{array}\right.
\end{eqnarray}That is,  let $a^{(0)}_{\mathcal{T},k_1}$ be a star if the row $\mathcal{T}$ of $\mathbf{B}^{(k_1)}$  is a star row, and be null otherwise.
\item {\bf Step 2.} Construction of $(\mathbf{A}^{(1)},\ldots,\mathbf{A}^{(K_1)})$. Note that there are in total $K_1Z_1=K_1{K_1K_2-K_2\choose t-K_2}$ star rows in $\mathbf{B}^{(1)}$, $\ldots$, $\mathbf{B}^{(K_1)}$ and each star row has $K_2$ star entries. Then we replace all these star entries in each star row of $\mathbf{B}^{(1)}$, $\ldots$, $\mathbf{B}^{(K_1)}$ with consecutive integers from $S+1$ to $S+K_2K_1{K_1K_2-K_2\choose t-K_2}$ to construct $(\mathbf{A}^{(1)},\ldots,\mathbf{A}^{(K_1)})$, and all these integers form the set $\mathcal{S}_{\text{M}}$ as follows.
    \begin{eqnarray}
    \label{eq-alphabet-m}
    \mathcal{S}_{\text{M}}=\left[S+1:\ S+K_2K_1{K_1K_2-K_2\choose t-K_2}\right]
    \end{eqnarray}
\item{\bf Step 3.} Construction of $\mathbf{A}$. We get an $F_1F_2\times (K_1+K_1K_2)$ array by arranging $\mathbf{A}^{(0)}$ and $(\mathbf{A}^{(1)}\ldots, \mathbf{A}^{(K_1)} )$ horizontally, i.e., $\mathbf{A}=\left(\mathbf{A}^{(0)}, \mathbf{A}^{(1)}, \ldots, \mathbf{A}^{(K_1)}\right)$.
\end{itemize}
Next we consider the integer sets of the constructed HPDA. $\mathcal{S}_{\text{M}}$ is given in \eqref{eq-alphabet-m}, and $|\mathcal{S}_{\text{M}}|=K_2K_1{K_1K_2-K_2\choose t-K_2}$. From the above construction, the integer set $\mathcal{S}_{k_1}$ of $\mathbf{A}^{(k_1)}$, $k_1\in[K_1]$ contains two parts. The first is the integer set of $\mathbf{B}^{(k_1)}$. Recall that $\phi_{t+1}(\cdot)$ is a bijection from  $\binom{[K_1K_2]}{t+1}$ to $[\binom{K_1K_2}{t+1}]$ in \eqref{Eqn_Def_AN}. Then for any sub-array $\mathbf{B}^{(k_1)}$, integer $s\in [S]$ is in $\mathbf{B}^{(k_1)}$ if and only if its inverse mapping $\mathcal{S}=\phi^{-1}_{t+1}(s)$ contains at least one integer of $[(k_1-1)K_2+1:k_1K_2]$, i.e., $\mathcal{S}\cap[(k_1-1)K_2+1:k_1K_2]\neq \emptyset$. So the integer set of $\mathbf{B}^{(k_1)}$ is
\begin{eqnarray}
\label{eq-S'_k_1}
\begin{split}
&\mathcal{S}'_{k_1}=\bigg\{\phi_{t+1}(\mathcal{S})\ |\mathcal{S}\cap[(k_1-1)K_2+1:k_1K_2]\neq\emptyset,
\mathcal{S}\in {[K_1K_2]\choose t+1}\bigg\},
\end{split}
\end{eqnarray}
where $|\mathcal{S}'_{k_1}|=S-{K_1K_2-K_2\choose t+1}$.
The remaining parts are integers used for replacing the star entries in the star rows of $\mathbf{B}^{(k_1)}$, thus the integer set of $\mathbf{A}^{(k_1)}$ is
\begin{eqnarray}
\label{eq-S_k_1}
\begin{split}
\mathcal{S}_{k_1}&=\left[S+(k_1-1)K_2{K_1K_2-K_2\choose t-K_2}+1:S+
k_1K_2{K_1K_2-K_2\choose t-K_2}\right] \bigcup \mathcal{S}'_{k_1}.&
\end{split}
\end{eqnarray}
Because the integers in $\mathcal{S}_{\text{M}}$ has no intersection with $\mathcal{S}'_{k_1}$, then we have $|\mathcal{S}_{k_1}| =$ $K_2{K_1K_2-K_2\choose t-K_2}+S-{K_1K_2-K_2\choose t+1}$.

\subsection{The verification of HPDA properties}
From \eqref{Eqn_Def_AN} the row $\mathcal{T}$ of $\mathbf{B}^{(k_1)}$ is a star row if and only if all the integers of $[(k_1-1)K_2+1:k_1K_2]$ are contained by $\mathcal{T}$. So there are ${K_1K_2-K_2\choose t-K_2}$ star rows, i.e., each column of $\mathbf{A}^{(0)}$ has $Z_1={K_1K_2-K_2\choose t-K_2}$ stars. Condition B$1$ of Definition \ref{def-H-PDA} holds.

Clearly $\mathbf{B}^{(k_1)}$ is a $(K_2,F,Z,|\mathcal{S}'_{k_1}|)$ PDA. Recall that $\mathbf{A}^{(k_1)}$ is obtained by replacing the star entries in star rows of $\mathbf{B}^{(k_1)}$ with some unique integers which have no intersection with $[S]$. Thus $\mathbf{A}^{(k_1)}$ is a $(K_2,F,Z_2,|\mathcal{S}_{k_1}|)$ PDA, satisfying Condition B$2$ of Definition \ref{def-H-PDA}.

From Step $2$, we know that each integer $s$ in $\mathcal{S}_{\text{M}}$ occurs only once in $(\mathbf{A}^{(1)},\ldots,\mathbf{A}^{(K_1)})$, then it's clear that the first part of Condition B$3$ holds. For any $a^{(k_1)}_{\mathcal{T},k_2}=s\in\mathcal{S}_{\text{M}}$, we have row $\mathcal{T}$ is a star row of $\mathbf{B}^{(k_1)}$, then by \eqref{eq-mirror-cache1} we have $a^{(0)}_{\mathcal{T},k_1}=*$. The second part of  Condition B$3$ holds, and Condition B$3$ holds.

We show that Condition B$4$ holds by the following reason. Assume that there are two entries $a^{(k_1)}_{\mathcal{T},k_2}=a^{(k^{'}_1)}_{\mathcal{T}^{'},k^{'}_2}=s$, where $k_1\neq k^{'}_1$. Then $s\not\in \mathcal{S}_{\text{M}}$ since each integer of $\mathcal{S}_{\text{M}}$ occurs exactly once. Furthermore if $a^{(k_1)}_{\mathcal{T'},k_2}=s'$ is an integer, then $s'$ must be the element of $\mathcal{S}_{\text{M}}$, otherwise it  contradicts our hypothesis that $\mathbf{B}$ is a PDA. From the construction of $\mathbf{A}^{(k_1)}$, $s'\in \mathcal{S}_{\text{M}}$ only if the row indexed by $\mathcal{T'}$ of $\mathbf{B}^{(k_1)}$ is a star row. Then by \eqref{eq-mirror-cache1} we have $a^{(0)}_{\mathcal{T'},k_1}=*$. Condition B$4$ of Definition \ref{def-H-PDA} holds.

From the above introduction, the constructed array is our expected $(K_1,K_2$; $F={K_1K_2\choose t}$; $Z_1={K_1K_2-K_2\choose t-K_2}$, $Z_2={K_1K_2-1\choose t-1}-{K_1K_2-K_2\choose t-K_2}$; $\mathcal{S}_{\text{M}},\mathcal{S}_1$, $\ldots$, $\mathcal{S}_{K_1})$ HPDA.
\subsection{Performance of The Related Scheme}
\label{cons1-perform}
According to Step $2$, the cardinality of integer set $\mathcal{S}_{\text{M}}\bigcap\mathcal{S}_{k_1}$ is actually the number of star entries in the star rows of $\mathbf{B}^{(k_1)}$, then we have $|\mathcal{S}_{\text{M}}\bigcap\mathcal{S}_{k_1}|=K_2{K_1K_2-K_2\choose t-K_2}$.
And the integer set $\bigcup_{k_1=1}^{K_1}\mathcal{S}_{k_1}\setminus\mathcal{S}_{\text{M}}$ is actually the integer set $[S]$, whose cardinality is ${K_1K_2\choose t+1}$. Based on the $(K_1,K_2$; $F={K_1K_2\choose t}$; $Z_1={K_1K_2-K_2\choose t-K_2}$, $Z_2={K_1K_2-1\choose t-1}-{K_1K_2-K_2\choose t-K_2}$; $\mathcal{S}_{\text{M}},\mathcal{S}_1$, $\ldots$, $\mathcal{S}_{K_1})$ HPDA and Theorem \ref{th-1}, we have an $F$-division $(K_1,K_2;M_1,M_2;N)$ coded caching scheme with security and privacy for a linear function retrieval problem whose performance is listed in \eqref{th-2-para}.

\section{Proof of Theorem \ref{th-3}}
\label{sec:proof-th-3}
In this section, we first describe how to construct an HPDA by the hybrid method based on two PDAs, and then we prove the obtained HPDA leads to a hierarchical coded caching scheme achieving the performance in Theorem \ref{th-3}.
\subsection{The Construction of Hybrid Method}
Given any $(K_1,F_1,Z_1,S_1)$ PDA $\mathbf{B}=(b_{f_1,k_1})_{f_1\in[F_1],\ k_1\in[K_1]}$ and $(K_2,F_2,Z_2,S_2)$ PDA $\mathbf{C}=(c_{f_2,k_2})_{f_2\in[F_2],\ k_2\in[K_2]}$, we show how to construct the $(K_1,K_2$; $F_1F_2$; $Z_1F_2$, $Z_2F_1$; $\mathcal{S}_{\text{M}}$, $\mathcal{S}_1$, $\ldots$, $\mathcal{S}_{K_1})$ HPDA $\mathbf{A}= \left(\mathbf{A}^{(0)}\right.$, $\mathbf{A}^{(1)}$, $\ldots$, $\left.\mathbf{A}^{(K_1)}\right)$ where
\begin{IEEEeqnarray}{rCl}
&& \mathbf{A}^{(0)}=(a^{(0)}_{(f_1,f_2),k_1}),\ f_1\in[F_1], f_2\in[F_2], k_1\in [K_1], \ \ \ a^{(0)}_{(f_1,f_2),k_1}\in\{*,null\}\nonumber,\\
&& \mathbf{A}^{(k_1)}=(a^{(k_1)}_{(f_1,f_2),k_2}),\ f_1\in[F_1], f_2\in[F_2], k_2\in [K_2],\ \ \  a^{(k_1)}_{(f_1,f_2),k_2}\in\{*\}\cup \mathcal{S}_{k_1},\   k_1\in [K_1]\nonumber.
\end{IEEEeqnarray} Note that we use the couple $(f_1,f_2)$ to indicate the $((f_1-1)F_1+f_2)$-th row of an array. The integer sets $\mathcal{S}_{\text{M}}$ and $\mathcal{S}_{k_1}$ are proposed in \eqref{eq-s_m} and \eqref{eq-S-k_1} respectively. The constructions of $\mathbf{A}^{(0)}$ and  $\left(\mathbf{A}^{(1)}\ldots, \mathbf{A}^{(K_1)} \right)$ are described  as  the following three steps:
\begin{itemize}
\item{\bf Step 1.} Construction of $\mathbf{A}^{(0)}$.
We can obtain an $F_1F_2\times K_1$ array $\mathbf{A}^{(0)}$ by deleting all the integers in $\mathbf{B}$ and replicating each row by $F_2$ times, which is the row size of the inner array in Step $2$. Then each entry of $\mathbf{A}^{(0)}$ can be written as follows.
\begin{eqnarray}
\label{eq-mirror-cache}
a^{(0)}_{(f_1,f_2),k_1}=\left\{\begin{array}{cc}
                  * & \text{if } b_{f_1,k_1}=*,\\
                  \text{null} & \text{otherwise}.
                \end{array}\right.
\end{eqnarray} 
\item{\bf Step 2.} Construction of $\left(\mathbf{A}^{(1)}\right.$$\ldots$,$\left.\mathbf{A}^{(K_1)}\right)$.
The main idea is replacing the entries of PDA $\mathbf{B}$ by inner array PDA $\mathbf{C}$ and adjusting the integers in inner arrays. As $\mathbf{B}$ consists of integer-type entries and star-type entries, our construction consists of the following two parts.
Firstly we replace each integer entry $b_{f_1,k_1}=s$ by an $F_2\times K_2$ array
\begin{eqnarray}
\label{eq-type-I}
\mathbf{I}_1(s)=\mathbf{C}+(s-1)\times S_2.
\end{eqnarray}
Secondly we replace each star entry $b_{f_1,k_1}=*$ by an $F_2\times K_2$ array
\begin{eqnarray}
\label{eq-type-II}
\mathbf{I}_2(k_1,f_1)=\mathbf{C}+[(k_1-1)Z_1+\varphi_{k_1}(f_1)-1+S_1]\times S_2,
\end{eqnarray} where $\varphi_{k_1}(f_1)$ represents the order of the row label $f_1$ from up to down among all the star entries in $k_1$-th column of $\mathbf{B}$. Then the $\left(\mathbf{A}^{(1)}\right.$$\ldots$,$\left.\mathbf{A}^{(K_1)}\right)$ is obtained. Here we take  $\left(\mathbf{A}^{(1)}\right.$, $\left.\mathbf{A}^{(2)} \right)$ in Fig. \ref{fig-sketch} as an example. As $b_{1,2}=b_{2,1}=s=1$, we have $\mathbf{I}_1(1)=$$\mathbf{C}+$$(1-1)\times3=\mathbf{C}$. While $b_{1,1}=b_{2,2}=*$, $\varphi_{1}(1)=\varphi_{2}(2)=1$, we have $\mathbf{I}_2(1,1)$ $=\mathbf{C}+[(1-1)\times1+1-1+1]\times3=\mathbf{C}+3$ and $\mathbf{I}_2(2,2)$ $=\mathbf{C}+[(2-1)\times1+1-1+1]\times3=\mathbf{C}+6$ in \eqref{eq-two-array}.
\begin{eqnarray}
\label{eq-two-array}
\mathbf{C}+3=\left(
\begin{array}{ccc}
*	&	4	&	5\\
4	&	*	&	6\\
5	&	6	&	*
\end{array}
\right)\ \ \
\mathbf{C}+6=\left(
\begin{array}{ccc}
*	&	7	&	8\\
7	&	*	&	9\\
8	&	9	&	*
\end{array}
\right)
\end{eqnarray}
In the above example, obviously the integer sets in $\mathbf{I}_1(1)$$=\mathbf{C}$, $\mathbf{I}_2(1,1)$$=\mathbf{C}+3$ and $\mathbf{I}_2(2,2)=\mathbf{C}+6$ have no common integer, and we can generalize the investigation to the general case, as illustrated in Lemma \ref{lem-integer-se-order}.

\begin{lemma}
\label{lem-integer-se-order}
For any integers $k_1$, $k_1'\in[K_1]$, $f_1$, $f_1'\in[F_1]$, $s$ and $s'$, we have the following statements on the integer sets in $\mathbf{I}_1(s)$, $\mathbf{I}_1(s')$, $\mathbf{I}_2(k_1,f_1)$ and $\mathbf{I}_2(k_1',f_1')$.
\begin{itemize}
\item 
    The integer sets in $\mathbf{I}_1(s)$ and $\mathbf{I}_1(s')$ have no common integer if and only if $s\neq s'$;
\item 
    When $k_1=k'_1$, the integer sets in $\mathbf{I}_2(k_1,f_1)$ and $\mathbf{I}_2(k'_1,f'_1)$ have no common integer if and only if $\varphi_{k_1}(f_1)\neq\varphi_{k'_1}(f'_1)$;
\item 
    When $k_1\neq k'_1$, the integer sets in $\mathbf{I}_2(k_1,f_1)$ and $\mathbf{I}_2(k'_1,f'_1)$ have no common integer;
\item 
    When $s\in[S_1]$, the integer sets in $\mathbf{I}_1(s)$ and $\mathbf{I}_2(k_1,f_1)$ have no common integer;
\end{itemize}
\end{lemma}
By Lemma \ref{lem-integer-se-order}, each entry $a^{(k_1)}_{(f_1,f_2),k_2}$, $k_1\in[K_1]$ in $\left(\mathbf{A}^{(1)}\right.$,$\ldots$,$\left.\mathbf{A}^{(K_1)}\right)$ is determined uniquely and can be written as follows,
\begin{IEEEeqnarray}{rCl}
\label{eq-user-cache}
a^{(k_1)}_{(f_1,f_2),k_2}=\left\{\begin{array}{l}
                  c_{f_2,k_2}\!+\!(s\!-\!1)S_2 ,\quad \text{if } b_{f_1,k_1}\!=\!s,\\
                  c_{f_2,k_2}\!+\!\big[(k_1\!-\!1)Z_1\!
                 + \varphi_{k_1}(f_1)\!-\!1\!+\!S_1\big]S_2 , \  \text{if } b_{f_1,k_1}=*.
                \end{array}\right.
\end{IEEEeqnarray}

\item{\bf Step 3.} Construction of $\mathbf{A}$.
\label{sec:con-hpda}
We get an $F_1F_2\times$$(K_1\!+\!K_1K_2)$ array by arranging $\mathbf{A}^{(0)}$ and $\left(\mathbf{A}^{(1)}\right.$,$\ldots$,$\left.\mathbf{A}^{(K_1)}\right)$ horizontally, i.e., $\mathbf{A}=\left(\mathbf{A}^{(0)}\right.$,$\mathbf{A}^{(1)}$,$ \ldots$,$\left. \mathbf{A}^{(K_1)}\right)$.
\end{itemize}
Next we consider the integer sets of the constructed HPDA. Firstly we consider $\mathcal{S}_{\text{M}}$, which is the union set of the integer set of each $\mathbf{I}_2(k_1,f_1)$, $k_1\in[K_1]$, $f_1\in[F_1]$. There are in total $K_1Z_1$ stars in $\mathbf{B}$, then by $\eqref{eq-type-II}$ and Lemma \ref{lem-integer-se-order} we have
\begin{eqnarray}
\label{eq-s_m}
\begin{split}
\mathcal{S}_{\text{M}}&=\left(S_1S_2:\ (S_1+Z_1K_1)S_2\right].&
\end{split}
\end{eqnarray}
Obviously the cardinality of $\mathcal{S}_{\text{M}}$ is $Z_1K_1S_2$. Secondly we focus on $\mathcal{S}_{k_1}$, i.e., the integer set of $\mathbf{A}^{(k_1)}$ for each $k_1\in[K_1]$. From Step 2, we know $\mathbf{A}^{(k_1)}$ is composed of $F_1-Z_1$ inner arrays $\mathbf{I}_1(s)$ and $Z_1$ inner arrays $\mathbf{I}_2(k_1,f_1)$. So the integer set $\mathcal{S}_{k_1}$ is actually the union set of all integer sets of the $F_1$ inner arrays. For a $(K,F,Z,S)$ PDA, we define $\mathcal{C}_{i}$ as the integer set containing all the integers in the $i$-th column where $\mathcal{C}_{i}\subset[S]$ and  $|\mathcal{C}_{i}|=F-Z$. Then by \eqref{eq-type-I}, \eqref{eq-type-II} and Lemma \ref{lem-integer-se-order},  the integer set of $\mathbf{A}^{(k_1)}$ is
\begin{eqnarray}
\label{eq-S-k_1}
\begin{split}
\mathcal{S}_{k_1}&=\left(((k_1-1)Z_1+S_1)S_2:(k_1Z_1+S_1)S_2\right]
\bigcup\left(\bigcup\limits_{s\in \mathcal{C}_{k_1},k_1\in [K_1] }\left(0+(s-1)S_2:sS_2\right]\right),
\end{split}
\end{eqnarray} $k_1\in [K_1]$.  Because $s\in[S_1]$, according to Lemma \ref{lem-integer-se-order}, all the $F_1$ integer sets of inner arrays do not have any common integer, so we have $|\mathcal{S}_{k_1}|=F_1S_2$.
\subsection{The Verification of HPDA Properties}
\label{subsub-verify}
Because there are $Z_1$ stars in each column of $\mathbf{B}$, from \eqref{eq-mirror-cache} each column of $\mathbf{A}^{(0)}$ has exactly $Z_1F_2$ stars, satisfying   Condition B$1$ of Definition \ref{def-H-PDA}.

Then we focus on Condition B$2$, i.e.,  $\mathbf{A}^{(k_1)}$ is a $(K_2$,$F_1F_2$,$F_1Z_2$,$ F_1S_2)$ PDA. Because $\mathbf{A}^{(k_1)}$ is composed of $F_1$ inner arrays, each column of $\mathbf{A}^{(k_1)}$ has $F_1Z_2$ stars. So Condition C$1$ of Definition \ref{def-PDA} holds. In the above we have $|\mathcal{S}_{k_1}|=F_1S_2$, obviously C$2$ of Definition \ref{def-PDA} holds.  Because all the arrays defined in \eqref{eq-type-I} and \eqref{eq-type-II} satisfy Condition C$3$ of Definition \ref{def-PDA} and by Lemma \ref{lem-integer-se-order}, the intersection of the integer sets of any two of the $F_1$ inner arrays is empty, then each $\mathbf{A}^{(k_1)}$ also satisfies the Condition C$3$. Thus,  each $\mathbf{A}^{(k_1)}$ is a $(K_2$,$F_1F_2$,$F_1Z_2$,$ F_1S_2)$ PDA. 

Next we  consider Condition B$3$. Recall that all the integers in $\mathcal{S}_{\text{M}}$ are generated from \eqref{eq-type-II}. If $k_1\neq k'_1$, by the third statement of Lemma \ref{lem-integer-se-order}, each integer in $\mathcal{S}_{\text{M}}$ only exists in one $\mathbf{A}^{(k_1)}$. When the entry  $a^{(k_1)}_{(f_1,f_2),k_2}=s\in \mathcal{S}_{\text{M}}$, from \eqref{eq-user-cache} we have $b_{f_1,k_1}=*$, and by \eqref{eq-mirror-cache} we have $a^{(0)}_{(f_1,f_2),k_1}=*$. Thus, Condition B$3$ holds.

Finally we consider the Condition B$4$. For any integers $k_1$, $k_1'\in [K_1]$, $k_2$, $k_2'\in [K_2]$ and any couples $(f_1,f_2)$, $(f_1',f_2')$, $f_1$, $f_1'\in [F_1]$, $f_2$, $f_2'\in [F_2]$ assume that $a^{(k_1)}_{(f_1,f_2),k_2}=
a^{(k'_1)}_{(f_1',f_2'),k'_2}=s$ is an integer. By Lemma \ref{lem-integer-se-order}, the case $k_1\neq k'_1$, $f_1=f'_1$ is impossible since the intersection of the integer sets of related inner arrays is empty. So we only need to consider the case where $k_1\neq k'_1$, $f_1\neq f'_1$, and there are three conditions:
\begin{itemize}
\item $a^{(k_1)}_{(f_1,f_2),k_2}$ and $a^{(k_1')}_{(f_1',f_2'),k_2'}$ are all in the arrays generated by \eqref{eq-type-II}. By the third statement of Lemma \ref{lem-integer-se-order}, this case is impossible since the intersection of the integer sets in related inner arrays is empty.
\item $a^{(k_1)}_{(f_1,f_2),k_2}$ and $a^{(k_1')}_{(f_1',f_2'),k_2'}$ are in the arrays generated by \eqref{eq-type-I} and \eqref{eq-type-II} respectively. By the forth statement of Lemma \ref{lem-integer-se-order}, this case is also impossible since the intersection of the integer sets in related inner arrays is empty.
\item $a^{(k_1)}_{(f_1,f_2),k_2}$ and $a^{(k_1')}_{(f_1',f_2'),k_2'}$ are all in the arrays generated by \eqref{eq-type-I}. Then we have $s=c_{f_2,k_2}+(s'-1)\times S_2=c_{f_2',k_2'}+(s''-1)\times S_2$ and it's     true if and only if $c_{f_2,k_2}=c_{f_2',k_2'}$ and $b_{f_1,k_1}=s'=b_{f'_1,k'_1}=s''$, because $c_{f_2,k_2}, c_{f_2',k_2'}\leq S_2$. Without loss of generality we assume that $a^{(k_1)}_{(f'_1,f'_2),k_2}$ is an integer entry. Because $b_{f_1,k_1}=b_{f'_1,k'_1}=s'$, and from Condition C$3$ of definition \ref{def-PDA} we have $b_{f_1',k_1}=b_{f_1,k_1'}=*$. According to \eqref{eq-mirror-cache} we have $a^{(0)}_{(f_1',f_2'),k_1}=*$.
\end{itemize}
 From the above discussion, the Condition B$4$ of Definition \ref{def-H-PDA} holds. Thus, $\mathbf{A}$ is our expected HPDA.
\subsection{Performance of the Related Scheme}
\label{cons2-perform}
According to Step $2$, for each $k_1\in[K_1]$, the integer set $\mathcal{S}_{\text{M}}\bigcap\mathcal{S}_{k_1}$ is the union set of the integer sets in $Z_1$ inner arrays $\mathbf{I}_2(k_1,f_1)$, $f_1\in[F_1]$, then we have $|\mathcal{S}_{\text{M}}\bigcap\mathcal{S}_{k_1}|=Z_1S_2$.
The integer set $\bigcup_{k_1=1}^{K_1}\mathcal{S}_{k_1}\setminus\mathcal{S}_{\text{M}}$ is the union set of  integer sets in $S_1$ inner arrays $\mathbf{I}_1(s)$, $s\in[S_1]$, then we have $|\bigcup_{k_1=1}^{K_1}\mathcal{S}_{k_1}\setminus\mathcal{S}_{\text{M}}|=S_1S_2$. Based on the $(K_1,K_2$; $F_1F_2$; $Z_1F_2$, $Z_2F_1$; $\mathcal{S}_{\text{M}}$, $\mathcal{S}_1$, $\ldots$, $\mathcal{S}_{K_1})$ HPDA and Theorem \ref{th-1}, we have an $F$-division $(K_1,K_2;M_1,M_2;N)$ coded caching scheme with security and privacy for a linear function retrieval problem whose performance is listed in \eqref{th-3-para}.

\section{Conclusion}
\label{sec:conclusion}
In this paper, we studied the hierarchical coded caching for the linear function retrieval problem and introduced a new combination structure, referred as HPDA, which can be used to characterize both the placement and delivery strategy. So the problem of designing a scheme for hierarchical network is transformed into constructing an appropriate HPDA. Then we showed by adding some random variables into the scheme derived from HPDA, we could obtain a secure and private scheme, where the files in the library and users' demands are protect against wiretapper who can overhear signals in any shared-link, and the users' demands are kept unknown to other colluding users and colluding mirror sites. Then we proposed a class of HPDAs by dividing the PDAs into several equal size groups, which leads to a class of schemes achieving the lower bound of the first layer transmission load $R_1$ for non-trivial cases. Due to the limitation of the system parameters in this class of HPDAs, we proposed another class of HPDAs via a hybrid construction of two PDAs. Consequently, using any two PDAs, a new HPDA can be obtained which   allows flexible system parameters and has a smaller subpacketization compared with our first class of HPDAs. In addition, the scheme derived from hybrid construction achieves a tradeoff between subpacketization and transmission load by choosing different PDAs.

\begin{appendices}
\section{Lower Bound of $R^*_1$ and $R_{1\_\text{linear}}^{*}$}
\label{appendix-optimal}
Assume that each user $\mathbf{U}_{k_1,k_2}$ requests a single file, denoted by $d_{k_1,k_2}$, $k_1\in[K_1]$, $k_2\in[K_2]$. Now we introduce an enhanced system where   each user already knows the cache contents of its connecting mirror site. For this enhanced system, denote the   cache contents of the user $\mathbf{U}_{k_1,k_2}$  as  $\bar{\mathcal{Z}}_{k_1,k_2}= \mathcal{Z}_{k_1}\cup \widetilde{\mathcal{Z}}_{k_1,k_2} $. In   uncoded placement scenario,      each file can be viewed as a collection of   $2^K$ packets as $W_{i}=\{W_{i,\mathcal{T}}|\mathcal{T}\subseteq[K_1]\times[K_2]\}$, where user  $\mathbf{U}_{k_1,k_2}$ stores $W_{i,\mathcal{T}} $ if $(k_1,k_2)\in\mathcal{T}$. Consider one permutation of $[K_1]\times[K_2]$ denoted by $\mathbf{u}=\{u_{1,1},u_{1,2},\ldots,u_{K_1,K_2}\}$, $u_{k_1,k_2}\in[K_1]\times[K_2]$, and one demand vector $\mathbf{d}=\{d_{1,1}, d_{1,2},\ldots,d_{K_1,K_2}\}$ where $d_{k_1,k_2}\neq d_{k'_1,k'_2}$ if $k_1\neq k'_1$ or $k_2\neq k'_2$.     We then   construct a genie-aided super-user with cached content
\begin{IEEEeqnarray}{rCl}
\bar{ {Z}}~&&=(\bar{\mathcal{Z}}_{u_{1,1}},\bar{\mathcal{Z}}_{u_{1,2}}\backslash(\bar{\mathcal{Z}}_{u_{1,1}}\cup W_{d_{u_{1,1}}}),\ldots,\nonumber\\
&&\quad\quad \bar{\mathcal{Z}}_{u_{K_1,K_2}}\backslash(\bar{\mathcal{Z}}_{u_{1,1}}\cup W_{d_{u_{1,1}}}\cup
\bar{\mathcal{Z}}_{u_{1,2}}\cup W_{d_{u_{1,2}}}\cup\cdots
\cup \bar{\mathcal{Z}}_{u_{K_1,K_2-1}}\cup W_{d_{u_{K_1,K_2-1}}}))).
\end{IEEEeqnarray}

The genie-aided super-user is able to recover $W_{d_{1,1}},$ $ W_{d_{1,2}},\ldots,$ $W_{d_{K_1,K_2}}$  from $(X,X_1,\ldots,X_{K_1},\bar{Z})$, where $X$ and $X_{k_1}$ are the signals sent by server and mirror site $\text{M}_{k_1}$, respectively. Thus, we have
 \begin{IEEEeqnarray}{rCl}
&& H(W_{d_{1,1}}, W_{d_{1,2}},\ldots,W_{d_{K_1,K_2}}|\bar{Z}) \nonumber\\
&&\quad  =  H(W_{d_{1,1}},\ldots,W_{d_{K_1,K_2}}|X,X_1,\ldots,X_{K_1},\bar{Z})
+I(W_{d_{1,1}}, \ldots,W_{d_{K_1,K_2}};X,X_1,\ldots,X_{K_1}|\bar{Z})\nonumber\\
&&\quad  =I(W_{d_{1,1}},\ldots,W_{d_{K_1,K_2}};X,X_1,\ldots,X_K|\bar{Z}) \leq H(X,X_1,\ldots,X_K|\bar{Z})\ {=}\  H(X|\bar{Z})\label{barZup}
\end{IEEEeqnarray}
 where  the last equality holds because $\bar{Z}$  contains all mirrors' contents $\mathcal{Z}_{1},\ldots,\mathcal{Z}_{K_1}$, leading to $H(X_{k_1}|X,\bar{Z})=H(X_{k_1}|X,\mathcal{Z}_{k_1})=0$ for all $k_1\in[K_1]$.

 Next we introduce a more powerful enhanced system    where   each user $\mathbf{U}_{k_1,k_2}$ has a caching size of $(M_1+M_2)B$ bits, and denote its cached content as   $\hat{\mathcal{Z}}_{k_1,k_2}$. Note that this enhanced system can only result in smaller communication loads in $R_1$ and $R_2$  than that of the first enhanced   system. This is because  in the new enhanced system  each users $\mathbf{U}_{k_1,k_2} $ is able to   cache any set of  sub-files of   $(M_1+M_2)B$ bits, including the caching strategy of the first enhanced system $\bar{\mathcal{Z}}_{k_1,k_2}= \mathcal{Z}_{k_1}\cup \widetilde{\mathcal{Z}}_{k_1,k_2} $. We then construct a new genie-aided super-user with cached content
\begin{IEEEeqnarray*}{rCl}\label{contentbhatZ}
\hat{ {Z}}~&&=(\hat{\mathcal{Z}}_{u_{1,1}},\hat{\mathcal{Z}}_{u_{1,2}}\backslash(\hat{\mathcal{Z}}_{u_{1,1}}\cup W_{d_{u_{1,1}}}),\ldots,\nonumber\\
&&\ \ \ \bar{\mathcal{Z}}_{u_{K_1,K_2}}\backslash(\hat{\mathcal{Z}}_{u_{1,1}}\cup W_{d_{u_{1,1}}}\cup \hat{\mathcal{Z}}_{u_{1,2}}\cup W_{d_{u_{1,2}}}\cup\cdots
\cup\hat{\mathcal{Z}}_{u_{K_1,K_2-1}}\cup W_{d_{u_{K_1,K_2-1}}}))).
\end{IEEEeqnarray*}
Due to the stronger caching ability of the new genie-aided super-user, we have
 \begin{IEEEeqnarray}{rCl}\label{keystepLower}
 H(W_{d_{1,1}}, W_{d_{1,2}},\ldots,W_{d_{K_1,K_2}}|\hat{Z})
 \leq  H(W_{d_{1,1}}, W_{d_{1,2}},\ldots,W_{d_{K_1,K_2}}|\bar{Z})
 \stackrel{(a)}{\leq} H(X|\bar{Z})\leq H(X).\quad
\end{IEEEeqnarray}
 where   (a) holds by \eqref{barZup}. From \eqref{keystepLower}, we obtain that
 \begin{IEEEeqnarray}{rCl}
 R_1\geq\sum_{(k_1,k_2)\in[K_1]\times[K_2]} \ \ \sum_{\mathcal{T}\subseteq[K_1]\times[K_2] \backslash\{u_{1,1},\ldots,u_{k_1,k_2}\}} \frac{W_{d_{u_{k_1,k_2}},\mathcal{T}}}{B}
 \label{Rlower1}
\end{IEEEeqnarray}
 This is equivalent to a single-layer coded caching system where the server connects with  $K_1K_2$-user each equipped with cache memory of $(M_1+M_2)B$ bits. Now we follow the method in \cite{Wang'ITW16} to prove the lower bound of $R^*_1$.

Summing all the inequalities in the form of \eqref{Rlower1} over all all permutations of
users and all demand vectors in which users have distinct demands, we obtain that
\begin{IEEEeqnarray}{rCl}
\label{Rlower2}
R_1\geq \sum_{t\in[0:K_1K_2]} \frac{\binom{K_1K_2}{t+1}}{N\binom{K_1K_2}{t}}x_t = \sum_{t\in[0:K_1K_2]} \frac{K_1K_2-t}{t+1} x_t
\end{IEEEeqnarray}
where $x_t=\sum_{i\in[N]}\ \ \sum_{\mathcal{T}\subseteq[K_1]\times[K_2]:|\mathcal{T}|=t} \frac{W_{i,\mathcal{T}}}{B}$. Also, we have the following conditions   due to the constraints on the file size and memory size
\begin{IEEEeqnarray}{rCl}\label{Rlower2Condition}
\sum_{t\in[0:K]} x_t =N,\quad  \sum_{t\in[0:K_1K_2]}tx_t=K_1K_2(M_1+M_2).
\end{IEEEeqnarray}
Combining \eqref{Rlower2} and \eqref{Rlower2Condition} and by Fourier Motzkin elimination, we obtain  the lower bound of $R_1^*\geq\frac{K_1K_2-t}{t+1}$ for $t\in[0:K_1K_2]$.
Note that the setting where each user requests a single file is included in linear function retrieval problem.  Then the optimal worst-case (the demand matrix is row full rank) load $R_{1\_\text{linear}}^{*}$ under the constraint of uncoded data placement is not decreased when the users request scalar linear functions of the files, so we have
\begin{eqnarray}
\label{eq-squee1}
  R_{1\_\text{linear}}^{*}\geq R_1^*\geq\frac{K_1K_2-t}{t+1}, ~\text{for~}  t\in[0:K_1K_2].
\end{eqnarray}
And by Lemma \ref{Lemma-optimality}, we have
\begin{eqnarray}
\label{eq-squee2}
  R_{1\_\text{linear}} = \frac{{K_1K_2-t}}{{t+1}}, ~\text{for~}  t\in[K_2:K_1K_2].
\end{eqnarray}
\end{appendices}
By combining \eqref{eq-squee1} and \eqref{eq-squee2}, we obtain
\begin{eqnarray}
\label{eq-squee}
  R_{1\_\text{linear}} = \frac{{K_1K_2-t}}{{t+1}}\geq R_{1\_\text{linear}}^{*}\geq R_1^*\geq\frac{K_1K_2-t}{t+1}, ~\text{for~}  t\in[K_2:K_1K_2].
\end{eqnarray}
By Squeeze Theorem and \eqref{eq-squee}, we obtain the lower bound
$R_{1\_\text{linear}}^{*}\!=\!\frac{K_1K_2-t}{t+1}, t\in[K_2:K_1K_2].$

\ifCLASSOPTIONcaptionsoff
  \newpage
\fi

\bibliographystyle{IEEEtran}
\bibliography{reference}

\end{document}